\documentclass[twocolumn]{aastex63}

\usepackage{amsmath}
\usepackage{caption}
\usepackage{subcaption}
\newcommand{\be}{\begin{equation}}
\newcommand{\ee}{\end{equation}}

\shorttitle{}
\shortauthors{}
\graphicspath{{./}{figures/}}

\begin{document}

\title{\Large{Periodic Fast Radio Bursts from Luminous X-ray Binaries}}

\correspondingauthor{Navin Sridhar}
\email{navin.sridhar@columbia.edu }

\author[0000-0002-5519-9550]{Navin Sridhar}
\affiliation{Department of Astronomy, Columbia University, New York, NY 10027, USA}
\affiliation{Columbia Astrophysics Laboratory, Columbia University, New York, NY 10027, USA}

\author[0000-0002-4670-7509]{Brian D. Metzger}
\affiliation{Center for Computational Astrophysics, Flatiron Institute, 162 W. 5th Avenue, New York, NY 10011, USA}
\affiliation{Columbia Astrophysics Laboratory, Columbia University, New York, NY 10027, USA}

\author[0000-0001-7833-1043]{Paz Beniamini}
\affiliation{Division of Physics, Mathematics and Astronomy, California Institute of Technology, Pasadena, CA 91125, USA}
\affiliation{Astrophysics Research Center of the Open University (ARCO), The Open University of Israel, P.O. Box 808, Ra’anana 43537, Israel}

\author[0000-0001-8405-2649]{Ben Margalit}
\altaffiliation{NASA Einstein Fellow}
\affiliation{Astronomy Department and Theoretical Astrophysics Center, University of California, Berkeley, Berkeley, CA 94720, USA}

\author[0000-0002-6718-9472]{Mathieu Renzo}
\affiliation{Columbia Astrophysics Laboratory, Columbia University, New York, NY 10027, USA}
\affiliation{Center for Computational Astrophysics, Flatiron Institute, 162 W. 5th Avenue, New York, NY 10011, USA}

\author[0000-0002-1227-2754]{Lorenzo Sironi}

\affiliation{Department of Astronomy, Columbia University, New York, NY 10027, USA}
\affiliation{Columbia Astrophysics Laboratory, Columbia University, New York, NY 10027, USA}

\author[0000-0003-3684-964X]{Konstantinos Kovlakas}

\affiliation{Geneva Observatory, University of Geneva, Chemin des Maillettes 51, 1290 
Versoix, Switzerland}
\affiliation{Physics Department, University of Crete, GR 71003, Heraklion, Greece}
\affiliation{Institute of Astrophysics, Foundation for Research and 
Technology-Hellas, GR 71110, Heraklion, Greece}



\begin{abstract}

The discovery of periodicity in the arrival times of the fast radio bursts (FRBs) poses a challenge to the oft-studied magnetar scenarios. However, models that postulate that FRBs result from magnetized shocks or magnetic reconnection in a relativistic outflow are not specific to magnetar engines; instead, they require only the impulsive injection of relativistic energy into a dense magnetized medium.  Motivated thus, we outline a new scenario in which FRBs are powered by short-lived relativistic outflows (``flares'') from accreting black holes or neutron stars, which propagate into the cavity of the pre-existing (``quiescent'') jet.  In order to reproduce FRB luminosities and rates, we are driven to consider binaries of stellar-mass compact objects undergoing super-Eddington mass-transfer, similar to ultraluminous X-ray (ULX) sources.  Indeed, the host galaxies of FRBs, and their spatial offsets within their hosts, show broad similarities with ULXs.  Periodicity on timescales of days to years could be attributed to precession (e.g., Lens-Thirring) of the polar accretion funnel, along which the FRB emission is geometrically and relativistically beamed, which sweeps across the observer line of sight. Accounting for the most luminous FRBs via accretion power may require a population of binaries undergoing brief-lived phases of unstable (dynamical-timescale) mass-transfer. This will lead to secular evolution in the properties of some repeating FRBs on timescales of months to years, followed by a transient optical/IR counterpart akin to a luminous red nova, or a more luminous accretion-powered optical/X-ray transient. We encourage targeted FRB searches of known ULX sources.

\end{abstract}


\section{Introduction}
\label{sec:intro}

Fast Radio Bursts (FRBs) are short, luminous pulses of coherent radio emission of extragalactic origin \citep{Lorimer+07, Keane+12, Thornton+13}. Among the many proposed FRB models \citep{Platts+19}, the best-studied are those that postulate flaring magnetars as their central-engines (e.g., \citealt{Popov&Postnov13,Kulkarni+14,Lyubarsky14,Katz16,Metzger+17,Beloborodov17,Kumar+17,Metzger+19,Wadiasingh2019}).  Magnetar models can account for many observed properties of FRBs, including their short (${\lesssim}$ millisecond) timescales, large energetics, potentially high polarization (e.g., \citealt{Michilli+18,Ruo+20}), ability to recur (e.g., \citealt{Spitler+16,CHIME_repeaters}), and frequent association with star-forming host galaxies \citep{Tendulkar+17,Bhandari+20,Heintz+20,Li&Zhang20,Mannings+20,Safarzadeh+20,Bochenek+20b}.  Magnetar models have also received support from the discovery of an FRB-like radio burst from the Galactic magnetar SGR1935+2154 (\citealt{CHIME+20,Bochenek+20}) in coincidence with a luminous, hard X-ray flare (e.g., \citealt{Mereghetti+20,Li+20}).

Despite these many successes, magnetar models are not without challenges:

\begin{itemize}
\item No confirmed magnetars (of the kind observed in our Galaxy and Local Group) are sufficiently active to explain the recurring FRB sources, such as FRB 121102 (e.g., \citealt{Spitler+16}) or the population of repeaters discovered by CHIME/FRB (e.g., \citealt{CHIME_repeaters}).  This is despite the fact that similarly active repeaters contribute a significant fraction of the total FRB rate, including the (currently) non-repeating population \citep{Margalit+20,Lu+20}.  Such behavior could be explained by invoking a younger and/or more active magnetar population, possibly with even stronger internal magnetic fields than those of known Galactic magnetars (e.g., \citealt{Beloborodov17}).  This required source population could even be the product of one or more rare formation channels (e.g., exotic supernovae, hereafter SNe, binary neutron star (NS) or NS-white dwarf mergers, or accretion-induced collapse of a white dwarf; \citealt{Metzger+17,Margalit+19,Zhong&Dai20,Margalit+20}), but these possibilities are currently speculative.

\item The two best-studied repeating sources, FRB 180916 and FRB 121102, show periodicities in their burst arrival times of ${\sim}$16 and ${\sim}$160 days, respectively \citep{CHIME+20b,Rajwade+20}.  Again, several plausible ideas were proposed to generate periodic behavior within magnetar and highly magnetized pulsar scenarios (precession, binarity, extremely slow rotation, orbital motion; \citealt{Lyutikov+20,Beniamini+20,Levin+20,Zanazzi&Lai20,Ioka&Zhang_20,Li&Zanazzi21}).  However, these would all be novel, as no confirmed magnetars exhibit these properties.

\item  Magnetar activity is ultimately limited by the energy contained in their strong magnetic fields.  This budget becomes strained once realistic radiative efficiencies for the FRB emission (e.g., \citealt{Plotnikov&Sironi19,Chen+20}) are combined with the long observed active lives of some repeating sources (approaching a decade for FRB 121102; \citealt{Cruces+21}).

\item One of the nearest repeating source, FRB 180916, is spatially offset in its host galaxy from the closet region of active star formation \citep{Tendulkar+21}.  While consistent with the demographics of larger FRB samples (e.g., \citealt{Mannings+20,Heintz+20}), this location is in tension with scenarios that invoke young (age ${\lesssim} 10$ kyr) magnetars formed in core collapse SNe \citep{Tendulkar+21}.

\end{itemize}

\begin{figure*}[t]
\fboxsep=0pt 
\fboxrule=1pt 
\includegraphics[width=0.95\textwidth]{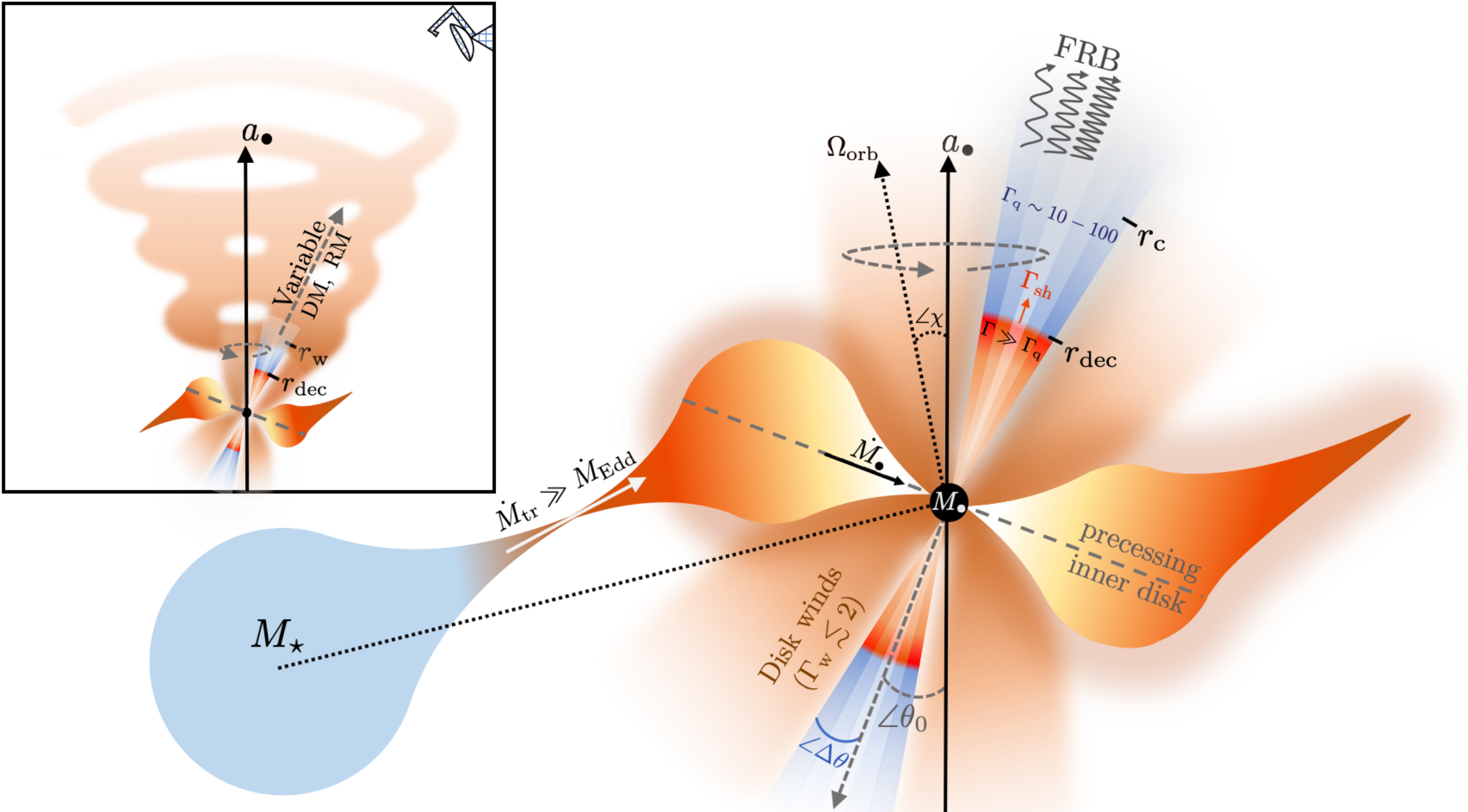}
\caption{Schematic diagram of the production of periodic FRBs from accretion-powered flares in ULX-like binaries outlined in this paper.  A star of mass $M_{\star}$ undergoes thermal- or dynamical-timescale mass-transfer onto a compact BH or NS remnant of mass $M_{\bullet}$ at rate $\dot{M}_{\rm tr}$ near or exceeding the Eddington rate.  The super-Eddington accretion flow is radiatively inefficient and hence subject to powerful outflows, which can reduce the accretion rate closer to the the central compact object, $\dot{M}_{\bullet} {\ll} \dot{M}_{\rm tr}$.  The mass-loaded disk winds also shape the narrow polar accretion funnel of half-opening angle $\Delta \theta$ and corresponding beaming fraction $f_{\rm b}$.  FRB emission is generated by a relativistic flare of energy released close to the innermost stable circular orbit (e.g. due to reconnection of magnetic field lines threading the BH horizon), which propagates outwards along the accretion funnel as an ultra-relativistic shock into the cavity of the previous ``quiescent'' jet.  Coherent radio emission (the FRB) is generated via the synchrotron maser shock mechanism or magnetic reconnection within the relativistic flow (at radii ${\gg} r_{\rm c}$, above which induced Compton scattering in the quiescent jet is negligible; eq.~\ref{eq:rc}).  If the spin axis of the BH is misaligned with respect to the angular momentum axis of the accretion disk, the polar cavity---and hence the direction along which the FRB emission is geometrically beamed---is modulated by Lens-Thirring (LT) precession on a timescale of days to years (eq.~\ref{eq:PLT}). \textit{Inset}: The magnetized disk outflows are swept into a spiral pattern due to the precession of the disk angular momentum about the axis of BH spin ($a_\bullet$). The instantaneous jet axis intersects this wind (from an earlier precession phase) on larger scales ${\gtrsim} r_{\rm w}$ (eq.~\ref{eq:rw}). Small systematic variations in the burst dispersion measure (DM) and rotation measure (RM) are expected due to this encounter (eq.~\ref{eq:RM}).}
\label{fig:cartoon}
\end{figure*}

Even within magnetar scenarios, there exist distinct models that invoke different physical mechanisms and emission regions for generating the FRB emission (e.g., \citealt{Zhang20b} for a review).  These can be broadly divided into ``magnetospheric'' scenarios (in which the emission originates close to the NS surface; \citealt{Kumar+17,Kumar&Bosnjak20,Wadiasingh2019,Wadiasingh2020}) and ``shock'' scenarios (in which the emission originates in a relativistic outflow at much greater distances; \citealt{Lyubarsky14,Beloborodov17,Plotnikov&Sironi19,Metzger+19}).

While magnetospheric scenarios may be exclusive to NS engines, shock scenarios apply to a wider range of central-engine models.  As emphasized by \citet{Metzger+19}, a basic requirement of the latter is simply the impulsive injection of ultra-relativistic energy into a strongly magnetized medium of the appropriate density (one example being the relativistic wind from the terminal stage of a binary NS merger; \citealt{Sridhar+21}).  Models in which FRBs are powered by magnetic reconnection in a relativistic magnetized outflow (e.g., \citealt{Lyubarsky20}) may be similarly agnostic to a magnetar as the specific engine triggering the energy release.

All of this motivates us to consider other central-engines, whose more continuous activity, longer timescales (e.g., binary orbit related), and greater energy reservoirs, concur more with the natural features of FRBs.  An obvious candidate explored in this paper are {\it accreting} stellar-mass compact objects, such as NSs or black holes (BHs), which are known to generate energetic and dynamical activity powered by strong magnetic fields carried in or generated by the accretion flow \citep[e.g.,][]{Fender+04}.  Accretion-powered FRB engines of this broad class were previously discussed by \citet{Waxman17} and \citet{Katz17, Katz20}.  As we shall demonstrate, the dual requirements to account for observed FRB timescales and luminosities drive us to the regime of super-Eddington accretion, for which the closest known observational analogs are the ``ultraluminous X-ray'' sources (ULXs) that are typically characterized by super-Eddington luminosities (see \citealt{Kaaret+17} for a review).  As we proceed, we shall therefore draw connections between the phenomenology of FRB and ULX sources. Fig.~\ref{fig:cartoon} is a schematic diagram of the envisioned scenario.

This paper is organized as follows.  Section \ref{sec:basic} places basic constraints on accretion scenarios and outlines a potential FRB mechanism.  Section \ref{sec:binary} discusses the types of binary systems that give rise to the requisite accretion rates and source formation rates.  Section \ref{sec:additional} outlines additional predictions of the scenario such as host galaxies and multiwavelength counterparts. We conclude by summarizing the predictions of the model in Section \ref{sec:conclusions}.

\section{Basic Constraints}
\label{sec:basic}

This section outlines the characteristics an accreting central-engine must satisfy in order to explain the basic observed properties of FRBs.  For this, we consider a compact object (fiducially a BH) of mass $M_{\bullet}$ accreting at a rate $\dot{M}_{\bullet}$, which we shall normalize to the Eddington accretion rate as $\dot{m} \equiv \dot{M}_{\bullet}/\dot{M}_{\rm Edd}$, where $\dot{M}_{\rm Edd} \equiv L_{\rm Edd}/c^{2}$, $L_{\rm Edd} = 4\pi GM_{\bullet}c/\kappa_{\rm es}$ is the Eddington luminosity, and we take $\kappa_{\rm es} = 0.38$ cm$^{2}$ g$^{-1}$ for the electron scattering opacity \citep{Frank+02}.

\subsection{Timescale}

The characteristic minimum timescale for significant energy release from a BH engine is set by the light-crossing time of the innermost stable circular orbit (ISCO):
\begin{equation}
    t_{\rm min} \sim \frac{R_{\rm isco}}{c} \approx 0.3\,{\rm ms}\,\,\left(\frac{m_{\bullet}}{10}\right)\left(\frac{R_{\rm isco}}{6R_{\rm g}}\right),
\label{eq:tmin}
\end{equation}
where $R_{\rm g} \equiv GM_{\bullet}/c^{2}$ and $m_\bullet \equiv M_\bullet/M_\odot$.  A similar expression holds in the case of an NS accretor with $R_{\rm isco}$ replaced by the NS radius, $R_{\rm NS} {\simeq} 12$ km (equivalent to $6R_{\rm g}$ for $M_{\bullet} = 1.4M_{\odot}$).

It is not obvious {\it a priori} that the timescale over which the engine releases energy would match that of an observed FRB.  However, this turns out to be approximately true in emission models that invoke dissipation within a transient relativistic outflow (Section \ref{sec:emission}).  Generic arguments show that FRBs must be produced in ultra-relativistic outflows (e.g.~\citealt{Lu&Kumar18}).  Consider a relativistically expanding ejecta shell, emitted as a ``flare'' from the central-engine over a timescale $t_{\rm f} {\gtrsim} t_{\rm min}$ with a corresponding thickness $\Delta {\simeq} c \cdot t_{\rm f}$, propagating toward the observer with a bulk Lorentz factor $\Gamma {\gg} 1$.  Any process that taps into the energy of the ejecta shell by crossing it at close to the speed of light in the comoving frame (e.g., a reverse shock or magnetic reconnection) will complete once the shell reaches a radius $r_{\rm FRB} \sim \Gamma^{2}\Delta$.  An FRB emitted coincident with this energy release will arrive to an external observer over a timescale ${\lesssim} r_{\rm FRB}/2\Gamma^{2} \sim t_{\rm f}$, i.e. roughly matching the original activity time of the central-engine\footnote{In FRB emission models powered by a shock propagating into a high magnetization $\sigma {\gg} 1$ medium, the observed emission duration can be shorter than that of the engine by a factor of $1/\sigma$ (e.g.~\citealt{Babul_Sironi_20,Beloborodov19}).} (see \citealt{Beniamini&Kumar20} for more details).  This follows a similar faithful mapping between engine and prompt emission activity in gamma-ray bursts (e.g., \citealt{Kumar&Zhang15}).

\subsection{Luminosity}

\begin{figure*}[t]
\begin{center}
\includegraphics[width=16.0cm]{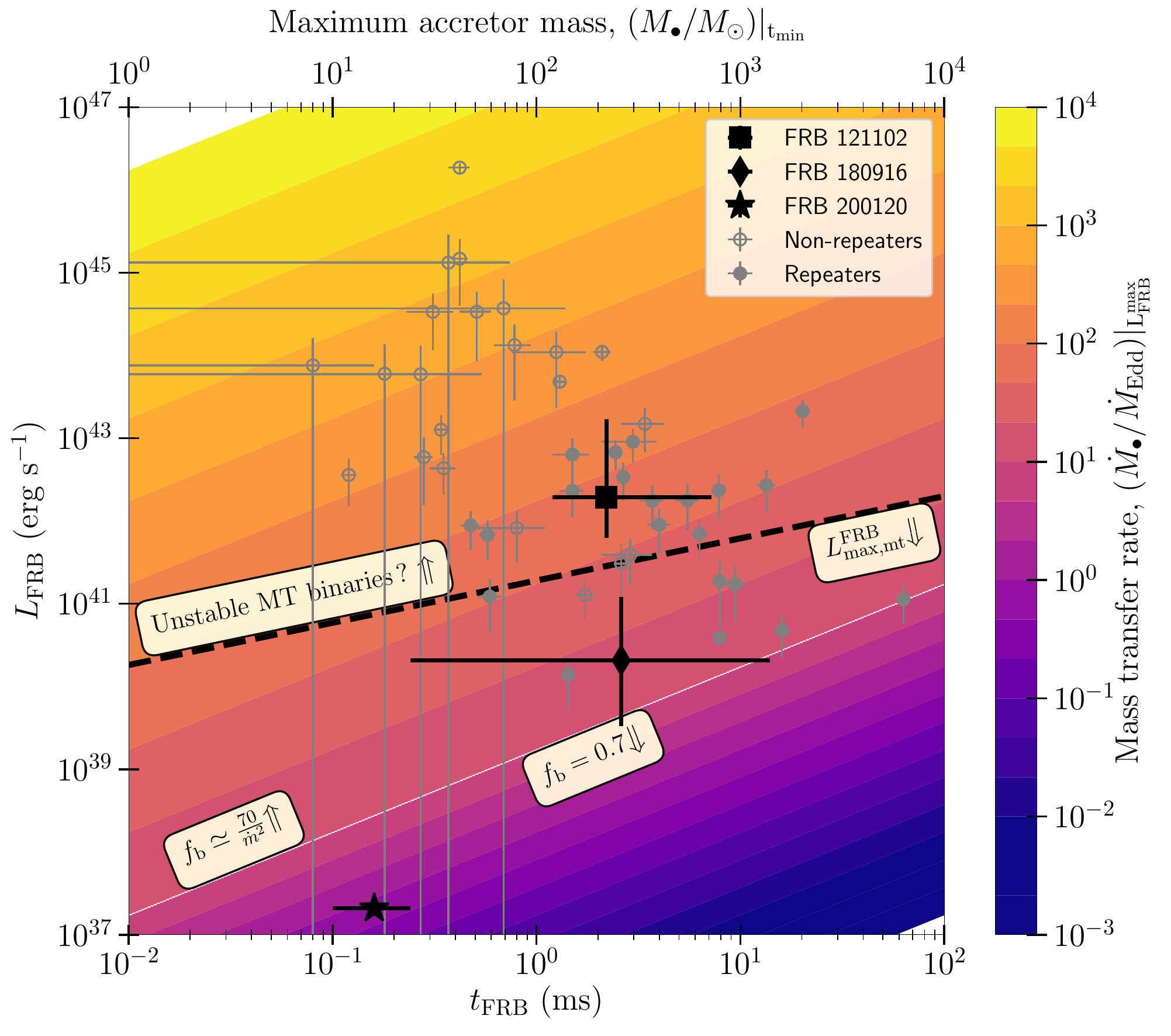}
\caption{If most of the luminous FRBs are powered by accreting compact objects, the latter must be NSs or stellar-mass BHs accreting near or well above the Eddington rate.  Here we show the observed isotropic luminosities and durations of repeating and non-repeating FRBs as filled and non-filled gray circles, including ranges for the repeating sources, FRB 121102 (square), FRB 180916 \citep[diamond; \url{http://www.frbcat.org};][]{Petroff+16} and FRB 200120 (star; \citealt{Bhardwaj+21}).  The top axis shows the NS/BH mass $M_{\bullet}$ corresponding to the minimum FRB duration $t_{\rm min}$ (eq.~\ref{eq:tmin}) assuming $R_{\rm isco} = 2 R_{\rm g}$.  Colored contours show the corresponding mass-transfer Eddington ratio required to achieve an isotropic-equivalent luminosity $L_{\rm FRB} = L_{\rm FRB}^{\rm max}$ (eq.~\ref{eq:Lmax}) for an assumed FRB emission efficiency $f_{\xi} = 10^{-3}$ (Section \ref{sec:emission}) using the value $M_{\bullet}$ corresponding to the $t_{\rm min}$ condition, and adopting a beaming fraction ($f_{\rm b}$) motivated by ULX observations (eq.~\ref{eq:fb}).  A dashed black line shows the limit corresponding to systems undergoing stable MT accretion (eq.~\ref{eq:Lmax2}).}
\label{fig:t-e}
\end{center}
\end{figure*}

The power of the ultra-relativistic component of a BH outflow is given by the \citet[BZ]{Blandford&Znajek77} jet luminosity, which can be written as \citep{Tchekhovskoy+11}
\be
L_{\rm BZ} = \eta \dot{M}_{\bullet}c^{2} \approx \eta \dot{m} L_{\rm Edd}, \label{eq:LBZ}
\ee
where the jet efficiency $\eta$ depends on the magnetic
flux threading the BH horizon and the dimensionless BH spin,
$a_{\bullet}$.  The maximum allowed $\eta = \eta_{\rm max}$
corresponds to a magnetically arrested accretion disk (MAD), and
varies from $\eta_{\rm max} =0.3$ (for $a_{\bullet} {\approx} 0.5$) to
$\eta_{\rm max} = 1.4$ \citep[for $a_{\bullet} = 0.99$;][]{Tchekhovskoy+11}.  A
similar jet luminosity can, in principle, be generated by an accreting
NS, although the magnetic flux in this case corresponds to that of the
NS surface field and the dependence of the jet power on accretion rate is more complicated (e.g., \citealt{Parfrey+16}).

Even absent a large-scale magnetic field of fixed polarity threading the disk, a temporary jet with a power $\sim L_{\rm BZ}$ can be generated by inflating and opening up small-scale field lines connecting the BH, due to the differential rotation of the accretion flow (e.g., \citealt{Parfrey+15,Ripperda+19,Mahlmann+20}).  The instantaneous jet efficiency can therefore in principle also approach $L_{\rm BZ}$ with $\eta \sim \eta_{\rm max} \sim 1$ for brief intervals even in relatively weakly magnetized disks. This differs from the MAD case insofar that the mean power corresponding to the persistent jet (when the flux from a reconnected magnetic loop is not threading the BH) may be significantly lower.  The ``transient'' nature of such a jet will turn out to be important in generating an observable FRB signal (Section \ref{sec:emission}).

The most luminous FRB sources will turn out to be rare compared to the super-Eddington accreting BH/NS population as a whole (Section \ref{sec:rates}), thus allowing them to represent the most extreme cases of accreting binaries found in nature (Section \ref{sec:rates}). Taking $\eta \sim \eta_{\rm max} \sim 1$ may be justified for their most luminous bursts. The maximum isotropic-equivalent radio luminosity of an FRB powered by accretion can thus be written
\begin{eqnarray}
L_{\rm FRB}^{\rm max} \approx f_{\xi}f_{\rm b}^{-1}\eta_{\rm max} \dot{M}_{\bullet}c^{2} \sim   f_{\xi}f_{\rm b}^{-1}\dot{m} L_{\rm Edd}  \nonumber \\
\underset{\dot{m} \gg 10}\approx 0.014\left(\frac{f_{\xi}}{10^{-3}}\right)\left(\frac{\dot{m}}{10}\right)^{3}L_{\rm Edd},
\label{eq:Lmax}
\end{eqnarray}
where $f_{\xi} \equiv L_{\rm FRB}/L_{\rm BZ}$ is the efficiency of converting the energy of the transient jet into coherent radio emission (e.g., $f_{\xi} < 10^{-2}$ in synchrotron maser scenarios; Section \ref{sec:emission}) and $f_{\rm b}$ is the beaming fraction.  The final line in eq.~\ref{eq:Lmax} employs an expression for the X-ray beaming fraction motivated by ULX observations (eq.~\ref{eq:fb}), an appropriate choice if the FRB is geometrically beamed by a plasmoid ejected along the accretion funnel (Section \ref{sec:beaming}).

Fig.~\ref{fig:t-e} shows the observed durations and luminosities of a large sample of FRBs, including the well-studied repeating sources FRB 121102 and FRB 180916, and the nearest extragalactic FRB 200120 \citep{Bhardwaj+21}.  The top axis gives the value of $M_{\bullet}$ corresponding to $t_{\rm min}$ (eq.~\ref{eq:tmin}) for $R_{\rm isco} {\simeq} 2 R_{\rm g}$ (a rough lower limit corresponding to a rapidly spinning BH).  Colored contours show the corresponding minimum Eddington ratio $\dot{m}$ required to achieve $L_{\rm FRB} = L_{\rm FRB}^{\rm max}$ (eq.~\ref{eq:Lmax}) for an assumed efficiency $f_{\xi} = 10^{-3}$ using the value $M_{\bullet}$ corresponding to the $t_{\rm min}$ condition (Section \ref{sec:emission}).  The contours are derived by assuming the beaming fraction motivated by ULX observations (eq.~\ref{eq:fb} and surrounding discussion).

Firstly, we see that observed durations of most require accretors of mass $M_{\bullet} {\lesssim}10^{3}M_{\odot}$, corresponding to NS or stellar-mass BH engines.  In order to explain the most luminous FRBs (particularly the non-repeating sources), we require super-Eddington accretion rates, $\dot{m} {\gtrsim} 1-10^{3}$. By contrast, the isotropic-equivalent luminosity of FRB 200120---purportedly located in the nearby galaxy M81---is smaller than any other extragalactic FRB by at least three orders of magnitude. Note that this luminosity is consistent with arising from an X-ray binary accreting below the Eddington rate.

\subsection{Beaming}
\label{sec:beaming}

Fig.~\ref{fig:t-e} demonstrates that the most luminous FRB sources require accretion rates $\dot{m} {\gg} 1$.  Such super-Eddington levels of accretion  \citep{Shakura&Sunyaev73} correspond to the regime of photon-trapped radiatively inefficient accretion flows (RIAFs; e.g., \citealt{Katz_1977,Abramowicz+88,Narayan&Yi95,Blandford&Begelman99}).  These systems are characterized by optically and geometrically thick accretion disks with powerful mass-loaded winds (e.g.~\citealt{Begelman+06,Poutanen+07}) and a narrow open funnel along the polar axis defined by the disk angular momentum.  Super-Eddington accretion is likely responsible for at least some of the ULX population, even when the accretor is a magnetized NS \citep{Bachetti+14,Mushtukov+15}.

To reproduce observed properties of ULXs, \citet{King09} estimated a geometric beaming fraction for the X-ray emission, corresponding to the opening solid angle of the accretion funnel, given by
\begin{align}
f_{\rm b,X} \approx
\Bigg\{\begin{array}{ll}
0.7,                    & \dot{m} \ll 10 \\
\frac{73}{\dot{m}^{2}}, & \dot{m} \gg 10
\end{array}
,
\label{eq:fb}
\end{align}
where the specific constant value of 0.7 in the $\dot{m} {\ll} 10$ limit is chosen for continuity with the $\dot{m} {\gg} 10$ limit.
Given the requirements of an extremely clean environment to generate an observable FRB  (Section \ref{sec:environment}), the relativistic outflow responsible for powering the FRB will itself likely be confined to within a similar solid angle (Fig.~\ref{fig:cartoon}).  In such a case, $f_{\rm b,X}$ provides at least a lower limit on the beaming fraction of the FRB itself, $f_{\rm b}$, which is defined as the fraction of $4\pi$ steradians into which the total power radiated by the FRB is channeled ($f_{\rm b}$ also denotes the ratio of the angle-integrated to isotropic-equivalent luminosity).

Section \ref{sec:periodicity} describes a scenario in which the periodic activity window of FRBs results from precession of the accretion disk's polar funnel (of half-opening angle $\Delta \theta$ and corresponding $f_{\rm b}$) across our line of sight (see also \citealt{Katz17}, who propose a similar geometry).  In this case, taking $f_{\rm b}=2\pi(1-\cos{(\Delta\theta)})/4\pi{\approx}\pi(\Delta\theta)^{2}/(4\pi)$, the duty cycle associated with the FRB activity window is then given by \citep{Katz21}:
\begin{align}
\zeta \approx   \frac{\Delta \theta}{\pi \theta_{0}} \approx \left(\frac{4f_{\rm b}}{\pi^{2}\theta_{0}^2}\right)^{1/2} \approx
\left\{\begin{array}{ll}
\frac{0.53}{\theta_0}, & \dot{m} \ll 10 \\
         \frac{5.4}{\dot{m}\theta_0}, 		                                            & \dot{m} \gg 10
                                       \end{array}
                               \right.,
\label{eq:zeta}
\end{align}
where $\theta_{0}$ is the angle of the axis of precession makes with respect to the observer's line of sight see Fig.~\ref{fig:cartoon}.  The observed activity window duty cycle $\zeta {\approx} 0.3-0.35$ of FRB 180916 \citep{CHIME+20b}, or $\zeta {\approx} 0.55$ for FRB 121102 \citep{Rajwade+20}, would then imply $\dot{m}\theta_0 {\lesssim} 10$ (Eq.\,\ref{eq:zeta} for $\dot{m}{\gg}10$).  For values $\theta_0 {\lesssim} 0.1-1$, the implied accretion rates $\dot{m} {\gtrsim} 10-100$ are broadly consistent with those required to power the FRB luminosities from these sources (Fig.~\ref{fig:t-e}).

\subsection {Clean Environment for FRB Escape}
\label{sec:environment}

A number of physical processes can absorb or attenuate radio waves in the vicinity of an accreting compact object.  One of the most severe is induced Compton scattering of the FRB beam by electrons surrounding the source (e.g., \citealt{Lyubarsky08}).  This makes it highly non-trivial to find the clean environment needed for the FRB to escape to an observer at infinity, given the dense outflows present in super-Eddington accretion systems.  The only plausible scenario is for the emission to be generated at large radii, far from the original launching point of the flare (inner magnetosphere), by relativistically expanding material directed outward along the narrow polar accretion funnel. (This funnel, prior to the FRB-generating flare, carried only the comparatively low-density jet, uncontaminated by the mass-loaded disk outflows present at wider angles.)

At the onset of the FRB-generating flare, the polar accretion funnel is filled by the plasma of whatever jet of lower power $L_{\rm q} = \eta_{\rm q}L_{\rm BZ}$ was present just prior to the flare, where here $\eta_{\rm q} {\lesssim} 1$ is the efficiency of the steady ``quiescent'' jet relative to that of the higher power $L_{\rm BZ} {\sim} \eta\dot{M}_{\bullet}c^{2}$ of the transient outburst responsible for powering the FRB (eq.~\ref{eq:LBZ}). Below, we discuss different physical interpretations of the low $\eta_{\rm q} {\ll} 1$ medium, such as if it truly indicates a less efficient jet (at the same accretion rate, $\dot{M}_{\bullet}$), or just a much lower $\dot{M}_{\bullet}$.

The lab-frame electron density of the quiescent jet at radius $r$ from the compact object is given by
\be
n_{\rm q} = \frac{\eta_{\rm q} L_{\rm BZ}}{4\pi (1+\sigma_{\rm q}) f_{\rm b} \Gamma_{\rm q} m_{\rm p} c^{3}r^{2}},
\label{eq:ne}
\ee
where $\sigma_{\rm q}$ and $\Gamma_{\rm q}$ are the magnetization and bulk Lorentz factor of the quiescent jet, respectively, and we have assumed an electron-ion jet composition\footnote{The composition of a BH or NS jet is likely to be dominated by electron/positron pairs on small scales close to the compact object (e.g. \citealt{Globus&Levinson13}).  However, by the larger radial scales of interest, the quiescent jet may have entrained baryons from the jet walls defined by the surrounding disk wind.}, where $m_{\rm p}$ is the proton mass.

The optical depth due to induced electron scattering above a radius $r$ in the wind, for an FRB of isotropic-equivalent luminosity $L_{\rm FRB}$, duration $t_{\rm FRB}$, and frequency $\nu_{\rm FRB}$, is approximately given by \citep{Lyubarsky08,Metzger+19}:
\begin{eqnarray}
\tau_{\rm c} \sim
\frac{3}{320 \pi^{2}}\frac{\sigma_{\rm T}}{m_{\rm e}}\frac{L_{\rm FRB} \cdot c t_{\rm FRB} \cdot n_{\rm q}}{\nu_{\rm FRB}^{3}r^{2}}\cdot \Gamma_{\rm q}^{3},
\label{eq:tauC}
\end{eqnarray}
where $\sigma_{\rm T}$ is the Thomson cross section, and the factor of $\Gamma_{\rm q}^{3}$ follows from the relativistic transformation of the induced scattering optical depth from the rest frame of the upstream wind \citep{Margalit+20}.\footnote{An additional complication arises from the impact of the strong wave of the FRB in accelerating the electrons in the upstream scattering medium to relativistic speeds.  However, \citet{Margalit+20} show that the enhancement in the optical depth due to this effect is canceled by the suppression of the scattering rate of the relativistic electrons \citep{Lyubarsky19}.}

Substituting eq.~\ref{eq:ne} with $L_{\rm q} = \eta_{\rm q}L_{\rm FRB}f_{\rm b}/f_{\xi}$ into eq.~\ref{eq:tauC}, we find that $\tau_{\rm c} {\lesssim} 1$ is achieved at radii satisfying
\begin{eqnarray}
r&\gtrsim&r_{\rm c} \approx \left[\frac{3 \sigma_{\rm T}\eta_{\rm q} L_{\rm FRB}^{2}t_{\rm FRB}\Gamma_{\rm q}^{2}}{1280 \pi^{3}m_{\rm e} m_{\rm p} c^{2}\nu_{\rm FRB}^{3}f_{\xi}}\right]^{1/4}\nonumber \\
 &\approx& 8\times 10^{13}{\rm cm}\,\frac{L_{40}^{1/2}t_{-3}^{1/4}\Gamma_{\rm q,2}^{1/2}}{\nu_{9}^{3/4}f_{\xi,-3}^{1/4}}\frac{\eta_{\rm q,-3}^{1/4}}{(1+\sigma_{\rm q})^{1/4}}, \label{eq:rc}
\end{eqnarray}
where $\eta_{\rm q,-3} \equiv \eta_{\rm q}/(10^{-3})$,
$\Gamma_{\rm q,2} = \Gamma_{\rm q}/100$,
$\nu_{9} = \nu_{\rm FRB}/10^{9}{\rm Hz}$,
$L_{40} \equiv L_{\rm FRB}/10^{40}{\rm erg\,s^{-1}}$,
$t_{-3} \equiv t_{\rm FRB}/1\,{\rm ms}$, $f_{\xi, -3} \equiv f_{\xi}/10^{-3}$.

To produce an FRB of duration $t_{\rm FRB} {\sim} r_{\rm FRB}/(2\Gamma^{2}c)$ at radius $r_{\rm FRB} > r_{\rm c}$, the Lorentz factor $\Gamma$ of the emitting region must obey
\be
\Gamma > \Gamma_{\rm min} \approx \left(\frac{r_{\rm c}}{2c t_{\rm FRB}}\right)^{1/2} \approx 400r_{c,13}^{1/2}t_{-3}^{-1/2}, \label{eq:Gammamin}
\ee
where $r_{\rm c,13} \equiv r_{\rm c}/(10^{13}{\rm cm})$, and we have assumed a mildly magnetized upstream $\sigma_{\rm q} {\lesssim} 1$.  The Thomson depth from the emission radius $r_{\rm FRB} > r_{\rm c}$ through the quiescent outflow at large radii is given by
\begin{eqnarray}
\tau_{\rm T} &\simeq& \int_{r_{\rm FRB}}^{\infty}n_{\rm q} \sigma_{\rm T}dr \approx \frac{\eta_{\rm q} L_{\rm BZ}\sigma_{\rm T}}{4\pi f_{\rm b} \Gamma_{\rm q} m_{\rm p} c^{3}r_{\rm FRB}(1+\sigma_{\rm q})}  \nonumber \\
&\approx& 1.5\times10^{-11}\frac{m_{\bullet}}{10}\frac{r_{\rm c}}{r_{\rm FRB}}\frac{\eta  \dot{m} }{r_{\rm c,13}f_{\rm b,-1}\Gamma_{\rm q,2}}\frac{ \eta_{\rm q,-3}}{(1+\sigma_{\rm q})},
\label{eq:tauT}
\end{eqnarray}
and the corresponding DM,

\begin{eqnarray} &&DM[{\rm pc\,cm^{-3}}] \simeq 5\times 10^{5}\frac{\tau_{\rm T}\Gamma_{\rm q}}{(1+z)} \nonumber \\
&\approx& \frac{10^{-3}}{(1+z)}\frac{r_{\rm c}}{r_{\rm FRB}} \frac{m_{\bullet}}{10}\frac{\eta \dot{m} }{r_{\rm c,13}f_{\rm b,-1}}\frac{\eta_{\rm q,-3} }{(1+\sigma_{\rm q})},
\label{eq:DM}
\end{eqnarray}
where $z$ is the source redshift, and the factor $\Gamma_{\rm q}$ arises from the Lorentz transformation of the measured DM.

In addition to requiring an FRB be freely able to propagate to Earth ($\tau_{\rm c} < 1$), one must also not overproduce the measured local DM value or its time variation between bursts from the same source, $\Delta {\rm DM}$ (for example, $\Delta {\rm DM} {\lesssim} 2$ pc cm$^{-3}$ in FRB 180916; \citealt{CHIME+20b}).
Eqs.~\ref{eq:rc}-\ref{eq:DM} reveal that satisfying both of these conditions is possible, but it requires (1) the FRB to be generated by a highly relativistic outflow, $\Gamma {\gg} 100$, (2) which propagates into the medium corresponding to a jet of extremely low luminosity, $\eta_{\rm q} {\ll} 10^{-2}$ and/or one with a very high magnetization $\sigma_{\rm q} {\gg} 1$.  The next section describes how similar requirements emerge within a specific FRB emission model.  Fig.~\ref{fig:quiescent} summarizes the allowed parameter space of quiescent jet properties needed to simultaneously satisfy constraints on the shock emission radius ($r_{\rm dec} > r_{\rm c}$) and on the lack of local time-variable DM contribution due to propagation through the jet.

Can accreting stellar-mass compact objects generate outflows with $\Gamma > 100$?  The jets of Galactic X-ray binaries achieve Lorentz factors of at least a few to tens (e.g., \citealt{Mirabel&Rodriguez94,Fender+04}), although the constraints are model dependent and generally amount to lower limits \citep{Miller-Jones+06}.  As in jets from active galactic nuclei (AGNs), radiative acceleration of optically thin gas to relativistic velocities is severely limited by radiation drag effects (e.g.~\citealt{Phinney82}); however, these effects are mitigated if the flow entrains enough matter to shield itself. From radiation acceleration alone, \citet{Begelman14} argue that jets from super-Eddington accretion systems can achieve $\Gamma {\simeq} \dot{m}^{1/4}$, corresponding to $\Gamma {\gtrsim} 5$ for $\dot{m} {\gtrsim} 10^{3}$. Even higher Lorentz factors can be achieved by acceleration resulting from magnetic dissipation of a high-$\sigma$ flow in an optically thick environment (e.g., \citealt{Drenkhahn&Spruit02}), and impulsively due to magnetic pressure gradients within the ejecta \citep{Granot+11}.  For example, an extreme limit is gamma-ray burst jets, which attain $\Gamma {\gtrsim} 100-1000$ (e.g.,~\citealt{Lithwick&Sari01}).  Taken together, it cannot be excluded that highly super-Eddington systems---unlike the majority of X-ray binaries in our Galaxy---are capable for brief periods of generating outflows that satisfy the constraints (\ref{eq:Gammamin}) and (\ref{eq:DM}).

We also require a low effective efficiency $\eta_{\rm q} {\ll} 1$, and/or large magnetization $\sigma_{\rm q} {\gg} 1$, in the quiescent jet.  A large contrast in jet power between the quiescent and an FRB-generating flaring state would be generated by a sudden (${\sim}$dynamical timescale) large increase in the mass accretion rate reaching the central object.  This could be achieved, for example, by a highly magnetized NS accretor transitioning between a state of steady accretion and a ``propeller'' regime in which accretion is prohibited (e.g.,~\citealt{Parfrey&Tchekhovskoy17}).  Although changes in the accretion rate up to a factor ${\sim} 10^{3}$ are indeed observed in Galactic X-ray pulsars \citep{Tsygankov+16}, the frequency of mode-switching would need to be much more rapid than observed to explain many recurring FRBs, which exhibit inter-burst intervals as short as ${\sim} 10-100$ s (e.g., \citealt{Gourdji+19}).  Large amplitude changes in the accretion rate (factor of ${\sim}10^3$) are also inferred in ULXs \citep{Kaaret&Feng09,Bachetti+14} and from the class of low-mass X-ray binaries known as soft X-ray transients (e.g., \citealt{Tanaka&Shibazaki96}), but again over timescales, much longer than needed to explain many FRB sources.

Even for a constant accretion rate $\dot{M}_{\bullet}$, an abrupt rise in the jet efficiency $\eta \propto \Phi_{\rm B}^{2}$ can be driven by an increase in the magnetic flux $\Phi_{\rm B}$ threading the BH horizon.  Although one has $\eta = \eta_{\rm q} = \mathcal{O}(1)$ for the jet of a spinning BH in a persistently MAD state \citep{Tchekhovskoy+10}, long phases of $\eta_{\rm q} {\ll} 1$ are possible if the inner disk only irregularly receives a large magnetic flux bundle \citep{Spruit&Uzdensky05,Parfrey+15}.

Finally, the above constraints are also satisfied even for a relatively constant jet power ($\eta_{\rm q} {\sim} 1$) if the quiescent jet is extremely highly magnetized $\sigma_{\rm q} {\gg} 1$ (and hence is low density) relative to the FRB-generating flare. However, in our calculations that follow, we focus on the case of moderate upstream magnetization and a large contrast between the flaring and quiescent jet luminosities.

\subsection{Emission Mechanism}
\label{sec:emission}

\begin{figure*}[t]
\begin{center}
\includegraphics[width=16.5cm]{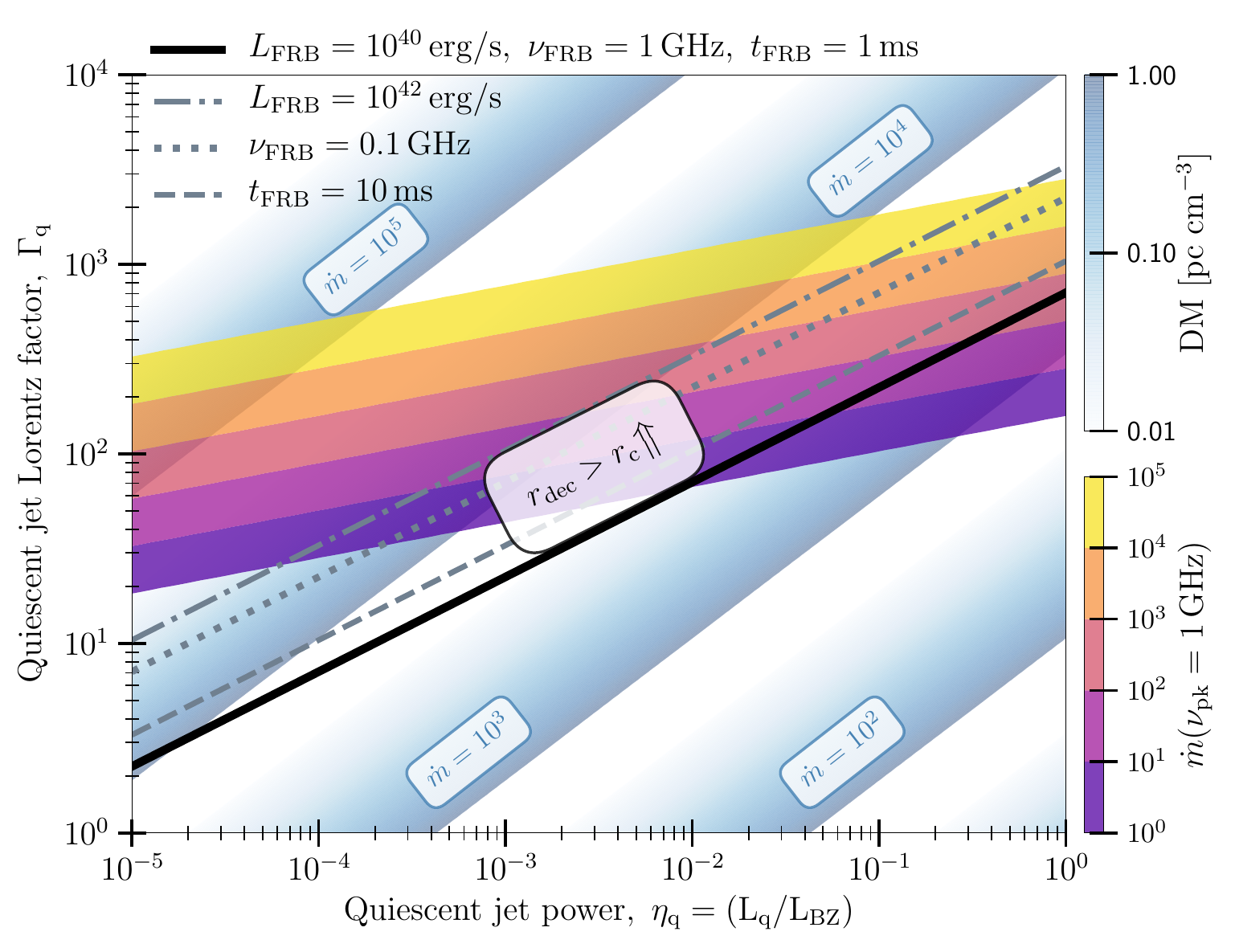}
\caption{The quiescent jet just prior to the FRB flare must be comparatively dilute and highly relativistic.  Here we show the allowed parameter space of the quiescent jet ($\eta_{\rm q}$, $\Gamma_{\rm q}$), based on the constraints from the optical depth for FRB escape, and on the local dispersion measure (DM) variation. The requirement of $r_{\rm dec}>r_{\rm c}$ for an optically thin upstream medium is demarcated by the diagonal lines (see eqs. \ref{eq:rc} and \ref{eq:rdec}). The solid black line corresponds to the ``fiducial FRB'' with $L_{\rm FRB}=10^{40}$\,erg\,s$^{-1}$, $\nu_{\rm FRB}=1$\,GHz, $t_{\rm FRB}=1$\,ms. The gray dash-dotted, dotted, and dashed lines denote the deviation of the burst luminosity ($L_{\rm FRB}=10^{42}$\,erg\,s$^{-1}$), frequency ($\nu_{\rm FRB}=0.1$\,GHz), and the duration ($t_{\rm FRB}=10$\,ms), respectively, from the fiducial case. The local environmental contribution to the DM, corresponding to different $\dot{m}\in[10,10^5]$, is denoted by the fading blue bands (assuming $r_{\rm FRB}=r_{\rm dec}$; see eqs. \ref{eq:DM} and \ref{eq:rdec}); the dark region corresponds to the expected upper limit of ${\sim}$1\,pc\,cm$^{-3}$. The values of $\eta_{\rm q}$ and $\Gamma_{\rm q}$ required to produce an emission peak at $\nu_{\rm pk}=1$\,GHz, for different $\dot{m}\in[1,10^5]$, are represented by the central yellow-violet ``emission contours'' (see eq.~\ref{eq:nupk}). For a given $\dot{m}$, the allowed range of $\eta_{\rm q}$ and $\Gamma_{\rm q}$ consists of the regions to the left of the intersection of the corresponding emission contour with the DM=1\,pc\,cm$^{-3}$ band, and above the $r_{\rm dec}=r_{\rm c}$ line. Throughout, we assume $\eta=1$, $\sigma_{\rm q}=1$, $f_{\xi}=10^{-3}$, flare duration $t_{\rm f}{\sim} t_{\rm min}=1$\,ms, redshift $z=0$, $f_{\rm b}{\approx}70/\dot{m}^2$, and $m_\bullet=10$.
}
\label{fig:quiescent}    
\end{center}
\end{figure*}

Advancing beyond the basic considerations of the previous section requires identifying a specific mechanism to convert a sudden release of relativistic energy from the engine (``flare'') into coherent radio emission at radii
$r_{\rm FRB} {\gg} r_{\rm c} {\sim} 10^{13}$ cm.  One such emission mechanism,
which we focus on for concreteness, is synchrotron maser emission from the relativistic shock generated as flare ejecta from the central-engine collide with a magnetized upstream  (\citealt{Lyubarsky14,Beloborodov17,Metzger+19}; see also \citealt{Waxman17}). In the present context, the upstream medium is the quiescent jet from earlier epochs.

Fast magnetosonic waves produced by plasmoid mergers in magnetic reconnection, which occur in the current sheets of BH magnetospheres (e.g.~\citealt{Philippov+19,Lyubarsky20}), or induced reconnection in the striped high-$\sigma_{\rm q}$ quiescent jet (and resulting inverse turbulent cascade; \citealt{Zrake+17}), provide an alternative emission mechanism for the flare.  However, the conditions for an FRB to escape from the magnetic reconnection-induced relativistically hot plasma are generally more severe than from a cold upstream \citep{Lyubarsky08}.

Returning to the shock scenario, we consider an accretion flare, which releases a transient ultra-relativistic outflow of power $L_{\rm f} {\sim} \eta \dot{M}_{\bullet}c^{2}$ (eq.~\ref{eq:LBZ}), duration $t_{\rm f} {\sim} t_{\rm min}$ (eq.~\ref{eq:tmin}), and radial bulk Lorentz factor $\Gamma_{\rm f} {\gg} \Gamma_{\rm min}, \Gamma_{\rm q}$.  Such flare ejecta could represent a magnetized plasmoid generated by a powerful reconnection event close to the BH/NS magnetosphere or light cylinder (e.g., \citealt{Parfrey+15,Ripperda+19,Yuan+20}).  The flare ejecta (with initial Lorentz factor $\Gamma$) will sweep up gas in the pre-existing quiescent jet (eq.~\ref{eq:ne}), generating a forward shock by a radius \citep{Sari&Piran95}
\be
r_{\rm dec} \approx 2\Gamma_{\rm sh}^{2}c\cdot t_{\rm f} \approx 2\times 10^{13}\eta_{\rm q,-3}^{-1/2}t_{\rm f,-3}\Gamma_{\rm q,2}^{2}\,{\rm cm},
\label{eq:rdec}
\ee
where \citep{Beloborodov17}
\be
\Gamma_{\rm sh} \simeq \Gamma_{\rm q}\left(\frac{L_{\rm f}}{L_{\rm q}}\right)^{1/4} \simeq \frac{\Gamma_{\rm q}}{\eta_{q}^{1/4}} \approx 560 \frac{\Gamma_{\rm q,2}}{\eta_{q,-3}^{1/4}},
\ee
 is the Lorentz factor of the shocked gas during the initial deceleration phase.  As discussed above, for the deceleration---and concomitant radio emission---to occur in an optically thin environment ($r_{\rm dec} {\gg} r_{\rm c}$; eq.~\ref{eq:rc}), we require a large quiescent jet Lorentz factor $\Gamma_{\rm q} {\gtrsim} 100$ and/or a large contrast $\eta_{\rm q} {\lesssim} 10^{-3}$, between the jet luminosity during the flare and that of prior quiescent state (Fig.~\ref{fig:quiescent}).

The synchrotron maser emission peaks at an electromagnetic frequency (e.g., \citealt{Gallant+92,Plotnikov&Sironi19})
\begin{eqnarray}
\nu_{\rm pk} &\sim& \frac{3\Gamma_{\rm sh} c}{2\pi r_{\rm L}} \sim \frac{3e\sigma_{\rm q}^{1/2} (L_{\rm f}L_{\rm q})^{1/4}}{2\pi m_{\rm e} c^{3/2}r_{\rm FRB}} \approx
\frac{e \eta^{1/2}\sigma_{\rm q}^{1/2}  L_{\rm Edd}^{1/2}}{4\pi m_{\rm e} c^{5/2}}\frac{r_{\rm dec}}{r_{\rm FRB}}\frac{\dot{m}^{1/2}\eta_{\rm q}^{3/4}}{t_{\rm f} \Gamma_{\rm q}^{2}} \nonumber \\
&\approx& 1.9\,{\rm GHz}\left(\frac{r_{\rm FRB}}{r_{\rm dec}}\right)^{-1}\left(\frac{m_\bullet}{10}\right)^{1/2}\left(\frac{\dot{m}}{10^{2}}\right)^{1/2}\frac{\eta^{1/2}\eta_{\rm q,-3}^{3/4}\sigma_{\rm q}^{1/2} }{\Gamma_{\rm q,2}^{2}t_{-3}},
\label{eq:nupk}
\end{eqnarray}
where $r_{\rm L} = \Gamma_{\rm q}m_{\rm e} c^{2}/e B_{\rm q}$ is the Larmor radius of electrons in the shocked plasma, $B_{\rm q} {\simeq} (L_{\rm q}\sigma_{\rm q}/cr_{\rm FRB}^{2})^{1/2}$ is the upstream lab-frame magnetic field of the quiescent jet material at the radius $r_{\rm FRB}$ of the FRB emission assuming $\sigma_{\rm q} {\gtrsim} 1$.  The efficiency of the FRB maser emission is $f_{\xi} {\sim} 10^{-3}$ for $\sigma_{\rm q} {\gtrsim} 1$ \citep{Plotnikov&Sironi19,Babul_Sironi_20}, motivating the choice for $L_{\rm FRB}^{\rm max}$ in Fig.~\ref{fig:t-e}.  Colored lines in Fig.~\ref{fig:quiescent} show the quiescent jet properties required to produce a burst of frequency $\nu_{\rm pk} {\sim} 1$ GHz for different values of $\dot{m}$.

As described previously in the context of other central-engine models, deceleration of the flare ejecta by the blast wave and its propagation to lower densities at radii ${\gtrsim} r_{\rm dec}$ produces downward drifting of $\nu_{\rm pk}$ and hence the frequency structure of the bursts \citep{Beloborodov19,Margalit+19,Metzger+19,Sridhar+21}, similar to that observed in the sub-bursts of FRB 121102 \citep{Hessels+19} and the CHIME repeaters \citep{CHIME_repeaters}\footnote{The downward drifting of $\nu_{\rm pk}$, and the burst sub-structures, are also possible for a radially decreasing upstream density profile \citep[e.g., the case of inspiral winds during NS mergers;][]{Sridhar+21}, and due to the non-linear propagation effects of FRBs \citep[e.g., self-modulation;][]{Sobacchi+21}, respectively.}.  For $\sigma_{\rm q} {\gg} 1$ the bursts are predicted to be nearly 100\% linearly polarized at the source, dominated by the so-called `X-mode' waves.  These waves have their electric field perpendicular to the upstream magnetic field, and hence also for a laminar quiescent jet that is parallel to the approximately fixed direction of the BH/NS spin vector. However, for lower $\sigma_{\rm q} {\lesssim} 1$, the shock also radiates an intense `O-mode', with power nearly comparable to the X-mode, which then deteriorates the polarization degree \citep{iwamoto_18}.  

A roughly constant polarization angle over many distinct bursts is consistent with observations of FRB 121102 (e.g.~\citealt{Michilli+18}) but in tension with other bursts that show polarization swings across the burst duration (e.g., \citealt{Ruo+20,Day+20}).  The latter require a more complicated magnetic field geometry in the upstream jet material than a laminar jet, for instance due to the effects of kink instabilities (e.g., \citealt{Bromberg&Tchekhovskoy16}) or the interaction of the jet with the accretion disk wind at large radii as a result of disk precession (Section \ref{sec:environment}).

\section{Binary Properties}
\label{sec:binary}

The majority of accreting Galactic NS/BH sources reside in long-lived X-ray binary systems with mass-transfer rates at or below the Eddington rate.  However, as shown in Fig.~\ref{fig:t-e}, such sources are energetically strained to explain most luminous FRBs. Nevertheless, there exist more extreme systems with much higher accretion rates. These include the microquasar binary SS433 \citep{Margon84,Fabrika04}, which is likely a BH being fed by a massive late A-type companion star at a super-Eddington rate \citep[e.g.,][]{hillwig:08}. Such a source viewed face-on would be a strong (and periodic due to jet precession) X-ray source akin to a ULX.\footnote{SS433 in fact exhibits X-ray emission with evidence for precession in the light curve \citep{Atapin&Fabrika16}, which are believed to arise from the jet rather than the accretion disk.}

One binary channel capable of generating sustained levels of highly super-Eddington accretion is stable thermal-timescale mass-transfer.  This can occur as an evolved massive secondary star undergoes Roche Lobe overflow, either on the main sequence or later when crossing the Hertzsprung gap to become a giant (e.g.,~\citealt{King&Begelman99,King+01,Rappaport+05, Wiktorowicz+15}).  As the stellar envelope becomes fully convective approaching the Hayashi track, the adiabatic response of the star to mass loss can lead to dynamically unstable mass-transfer, which manifests as a ``common envelope event'' engulfing the system in gas and ultimately precluding the clean environment necessary for FRB formation.  However, as emphasized by \citet{Marchant+17} and \citet{Pavlovskii+17}, a large fraction of systems nominally in the unstable regime may in fact undergo stable mass-transfer due to their outermost surface layers remaining radiative.\footnote{Making such systems stable also reduces the predicted rate of binary BH mergers from population synthesis models, bringing them into better accord with observations by LIGO/Virgo.}

This section considers whether such systems provide the accreting BH/NS systems required to power FRBs as discussed thus far.  We initially focus on stable mass-transfer systems, and, finding it possibly insufficient to account for the most luminous FRBs, we return to the shorter-lived unstable systems at the end.

\subsection{Mass-transfer Rate}
\label{sec:transfer}

For thermal-timescale mass-transfer from a star of mass $M_\star = m_{\star} M_{\odot}$ in a semi-detached binary to the companion of mass $m_{\bullet}$ (in our case, a BH or NS), the mass-transfer rate can achieve a value
(e.g., \citealt{Kolb98}), \be \dot{M}_{\rm tr} \sim \frac{M_\star}{\tau_{\rm
    KH}} \sim 3\times 10^{-8}m_\star^{2.6}M_{\odot}{\rm
  yr^{-1}}, \label{eq:Mdottr} \ee where \be \tau_{\rm KH} \approx
3\times 10^{7}\frac{m_\star^2}{r_{\star}l_{\star}}{\rm yr} \approx
3\times 10^{7}m_\star^{-1.6}{\rm yr}
\label{eq:tKH}
\ee is the Kelvin-Helmholtz time of the star when it leaves the main
sequence and $L_{\star} = l_\star L_{\odot}$ is its luminosity.  In
the final equalities we have assumed a main sequence star of mass $1 {\lesssim} m_\star {\lesssim} 40$ with a radiative envelope, for which $r_\star \propto m_\star^{0.6}$ and
$l_\star \propto m_\star^{3}$.

On the other hand, $\dot{M}_{\rm tr}$ cannot be arbitrarily large, or the super-Eddington disk cannot ``fit'' into the binary.  More precisely, if the disk is locally super-Eddington even at the circularization radius, then a thick disk encompasses the entire system and common envelope-like runaway will still occur.  \citet{King&Begelman99} estimated this occurs for $m_\star {\gtrsim} 53 m_\bullet^{0.18}$, placing an Eddington-scaled upper limit on $\dot{M}_{\rm tr}$ of
\be \frac{\dot{M}_{\rm tr}^{\rm max}}{\dot{M}_{\rm Edd}} \sim 1.2\times
10^{5}\left(\frac{m_{\bullet}}{10}\right)^{-0.53}.
\label{eq:mdottrmax}
\ee 

To power FRB emission, we are interested in the accretion rate reaching the innermost radii of the disk.  However, for super-Eddington accretion, only a small fraction of the transferred mass is expected to reach the BH due to massive winds (e.g.~\citealt{Blandford&Begelman99}).  In particular, for $\dot{M}_{\rm tr} {\lesssim} \dot{M}_{\rm tr}^{\rm max}$ we expect 
\be
\dot{M}_{\bullet} \simeq \dot{M}_{\rm tr}\left(\frac{R_{\rm
      Edd}}{R_{\rm isco}}\right)^{-p},
\label{eq:ADAF}
\ee
where
$R_{\rm Edd} {\simeq} R_{\rm g}\left(\frac{\dot{M}_{\rm tr}}{\dot{M}_{\rm Edd}}\right)$ is the trapping radius interior to which the disk becomes locally super-Eddington and $0<p<1$.  Taking a value $p {\approx} 0.7$ motivated by hydrodynamical simulations of RIAFs (\citealt{Yuan&Narayan14}), we obtain
\be
\dot{m} = \left(\frac{R_{\rm isco}}{R_{\rm g}}\right)^{0.7}\left(\frac{\dot{M}_{\rm tr}}{\dot{M}_{\rm Edd}}\right)^{0.3}.
\label{eq:mdotRIAF}
\ee
The upper limit on the mass-transfer rate (\ref{eq:mdottrmax}) then becomes an upper limit on the accretion rate reaching the central compact object,
\be
\dot{m} \lesssim 54\left(\frac{R_{\rm isco}}{2R_{\rm g}}\right)^{0.7}\left(\frac{m_{\bullet}}{10}\right)^{-0.16}
\label{eq:mdotmax}
\ee
From eq.~\ref{eq:Lmax}, we obtain a theoretical maximum of the isotropic-equivalent FRB luminosity for stably accreting sources:
\be
L_{\rm FRB,mt}^{\rm max} \approx 2.2 L_{\rm Edd}
 f_{\xi,-3}\left(\frac{R_{\rm isco}}{2 R_{\rm g}}\right)^{2.1}\left(\frac{m_{\bullet}}{10}\right)^{-0.53},
 \label{eq:Lmax2}
\ee
where we have taken $f_{\rm b} = f_{\rm b,X}$ (eq.~\ref{eq:fb}) in the $\dot{m} {\gg} 10$ limit (as satisfied for $m_{\bullet} {\lesssim} 10^{3}$ for $R_{\rm isco} {\sim} R_{\rm g}$).

The constraint (\ref{eq:Lmax2}) is shown as a black line in Fig.~\ref{fig:t-e}.  Many observed FRBs can in principle satisfy the $L_{\rm FRB,mt}^{\rm max}$ limit.  However, it is violated by the most luminous sources, particularly many (currently) non-repeating sources.  These FRBs could still be powered by accreting systems if the latter are undergoing {\it unstable} mass-transfer at a rate $\dot{M}_{\rm tr} {\gg} \dot{M}_{\rm tr, max}$.  Such systems have just begun mass-transfer but are in the process of evolving toward a merger or common envelope \citep[e.g.,][]{macleod:20}.

The lifetime of unstable systems as FRB sources corresponds to the timescale for runaway accretion.  Based on observations and modeling of the stellar merger V1309 Sco \citep{Tylenda+11,Pejcha14,Pejcha+17} this evolution time can be estimated as\footnote{In the luminous red nova AT 2018bwo, \citet{Blagorodnova+21} inferred that a state of thermal-timescale unstable mass-transfer was maintained for nearly a decade prior to evolving toward a common envelope.} 
\be \tau_{\rm unst} \sim (5-100)P_{\rm orb} \sim (10-10^5){\rm d},
\label{eq:trun}
\ee 
where $P_{\rm orb} {\sim} 1-1000$ {\rm d} is the binary orbital period (eq.~\ref{eq:Porb}).  The resulting mass-transfer rate, $\dot{M}_{\rm tr} {\sim} M_{\star}/\tau_{\rm unst} {\sim} 10^{3}-10^{7}\dot{M}_{\rm Edd}$ (e.g., for $M_{\star} {\sim} 10M_{\odot}$), can in principle exceed $\dot{M}_{\rm tr}^{\rm max}$ (eq.~\ref{eq:mdottrmax}) by orders of magnitude.  Even with accounting for mass-loss from disk winds this results in accretion rates reaching the compact object $\dot{m} {\gtrsim} 100$ (eq.~\ref{eq:mdotRIAF}), sufficient to explain the most luminous FRB sources (Fig.~\ref{fig:t-e}).  As the merger proceeds, the accretion rate rises exponentially, and the nature of the gaseous environment surrounding the binary will become increasingly ``messy.''  This will eventually lead to a cessation of FRB activity, at the very latest once the compact object plunges into the donor star or is completely engulfed by the common envelope on the timescale ${\sim} \tau_{\rm unst}$.

If a system undergoing unstable accretion were to generate multiple FRBs, the ``runaway'' process should impart a systematic variation in the FRB properties approaching the dynamical plunge.  For example, as the accretion rate rises---and the opening angle of the accretion funnel shrinks---the isotropic-equivalent luminosity of the bursts could increase in time, at least initially.

\subsection{Source Rates}
\label{sec:rates}

Under the assumption that all FRBs repeat with a luminosity function similar to the repeat bursts from FRB 121102, \citet{Nicholl+17} placed a constraint on the volumetric space density of FRB sources, $\mathcal{N}_{\rm FRB}$ (see also \citealt{Lu&Kumar16}).  The latter can expressed as the product of the FRB source formation rate $\mathcal{R}$ and the average active lifetime $\tau$, viz.~
\be
\mathcal{N}_{\rm FRB} = \mathcal{R}\cdot \tau \sim 1.3\times 10^{4}\left(\frac{f_{\rm b}}{0.1}\right)^{-1}\left(\frac{\zeta}{0.1}\right)^{-1}{\rm Gpc^{-3}}, \label{eq:NFRB}
\ee
where $f_{\rm b}$ and $\zeta$ are the average beaming fraction of each FRB and the duty cycle of the active window, respectively (which may be related by the common geometry of the accretion funnel; Section \ref{sec:periodicity}).

One relevant comparison is to the volumetric rate of ULXs sources.  The local rate of all ULXs (defined by X-ray luminosities $L_{\rm X} > 10^{39}$ erg s$^{-1}$) is $\mathcal{N}_{\rm ULX,39} {\sim} 2\times 10^{7}$ Gpc$^{-3}$ \citep{Swartz+11}.  However, the luminosity function steeply decreases moving to the higher isotropic-equivalent luminosities ($>L_{\rm X} {\sim} 10^{40}$ erg s$^{-1}$) required in our scenario to power FRBs, exhibiting a sharp break above $L_{\rm X} {\sim} 10^{40}$ erg s$^{-1}$, such that the rate above $L_{\rm X} {\sim} 10^{41}$ erg s$^{-1}$ is $\mathcal{N}_{\rm ULX,41}  {\sim} 10^{4}$  Gpc$^{-3}$ \citep{Mineo+12}.  Yet more luminous ULXs with $L_{\rm X} {\gtrsim} 10^{42}$ erg s$^{-1}$, such as the ``hyper-luminous'' source HLX-1 \citep{Farrell+09}, are even less common \citep{Gao+03}.

For stably accreting systems, the maximum FRB active lifetime is of the
order of the Kelvin-Helmholtz time, $\tau_{\rm KH}$ (eq.~\ref{eq:tKH}).  Accounting for the bulk of the FRB population through this channel ($\tau = \tau_{\rm KH}$ in eq.~\ref{eq:NFRB}) then requires a source formation rate
\be
\mathcal{R}_{\rm stable} \sim 0.02\left(\frac{m_{\star}}{10}\right)^{1.6}\left(\frac{f_{\rm b}}{0.1}\right)^{-1}\left(\frac{\zeta}{0.1}\right)^{-1}{\rm Gpc^{-3}\,yr^{-1}}.
\ee
\citet{Pavlovskii+17} estimate a Milky Way formation rate of ${\sim} 3\times 10^{-5}$ yr$^{-1}$ binaries with mass-transfer rates $\dot{M}_{\rm tr} {\gtrsim} 100\dot{M}_{\rm Edd}$, corresponding roughly to $\mathcal{R} {\sim} 100$ Gpc$^{-3}$ yr$^{-1}$.  The rate is higher in sub-solar metallicity environments (Section \ref{sec:hosts}).  Thus, only a small fraction ${\lesssim} 10^{-3}$ of potentially stable super-Eddington systems need to serve as active FRB sources.  This suggests an extra variable would need to be responsible for making a small fraction of ULX binaries into FRB sources.

Several special conditions may indeed need to be met to produce an observable FRB (Section \ref{sec:emission}). For example, generating a flare of sufficient power may require a large magnetic flux threading the BH, a task aided if the mass-transferring star is itself highly magnetized.  Although strongly magnetized massive stars are relatively common (e.g., the magnetic A-stars, with a 10\% occurrence rate), very few magnetic stars are in binaries (see \citealt{Shultz+15} for
an exception).  This low binary fraction could be explained if strong magnetic fields are the result of stellar
mergers (e.g. \citealt{Schneider+19}).  In such a case, forming a mass-transferring binary with a magnetized secondary might require a low-probability scenario, such as chaotic evolution or Kozai-Lidov oscillations in an initial hierarchical triple or quadrupole system (which acts to bring a magnetized stellar merger product into contact with the NS/BH primary).

The most luminous FRB sources require the high accretion rates achieved by binary systems undergoing unstable mass-transfer just prior to merging or entering a common envelope phase.  Applying the shorter lifetime of unstable systems, $\tau_{\rm unst} {\ll} \tau_{\rm KH}$ (eq.~\ref{eq:trun}) to eq.~\ref{eq:NFRB}, leads to a higher required rate of unstable events:
\begin{eqnarray} \mathcal{R}_{\rm unst} &\sim& 5\times 10^3\left(\frac{f_{\rm unst}}{0.1}\right)\left(\frac{\tau_{\rm unst}}{100P_{\rm orb}}\right)^{-1}\left(\frac{P_{\rm orb}}{10\,{\rm d}}\right)^{-1} \nonumber \\
&\times& \left(\frac{f_{\rm b}}{0.01}\right)^{-1}\left(\frac{\zeta}{0.1}\right)^{-1}{\rm Gpc^{-3}\,yr^{-1}},
\label{eq:Runstable}
\end{eqnarray}
where $f_{\rm unst}$ is the fraction of FRBs that arise from unstably accreting systems, and we have scaled $f_{\rm b}$ to the smaller value expected for $\dot{m} {\gg} 10$ (eq.~\ref{eq:fb}).

It is useful to compare the rate of eq.~\ref{eq:Runstable} to that of luminous red nov\ae{} (LRNe), optical transients from stellar mergers triggered by unstable mass-transfer (\citealt{Soker&Tylenda06,
Tylenda+11}).  The most luminous LRNe, which arise from the merger of massive stars ${\gtrsim} 10M_{\odot}$, occur at a rate of $\mathcal{R}_{\rm LRNe} {\sim} 10^{5}$ Gpc$^{-3}$ yr$^{-1}$ \citep{Kochanek+14}.  Although the merging systems of harboring BH/NS primaries of interest are even rarer than ordinary massive star mergers, their rates may still be consistent with the most luminous FRBs arising from unstable binaries.

\subsection{Periodicity}
\label{sec:periodicity}

\begin{figure}[t]
\includegraphics[width=8cm]{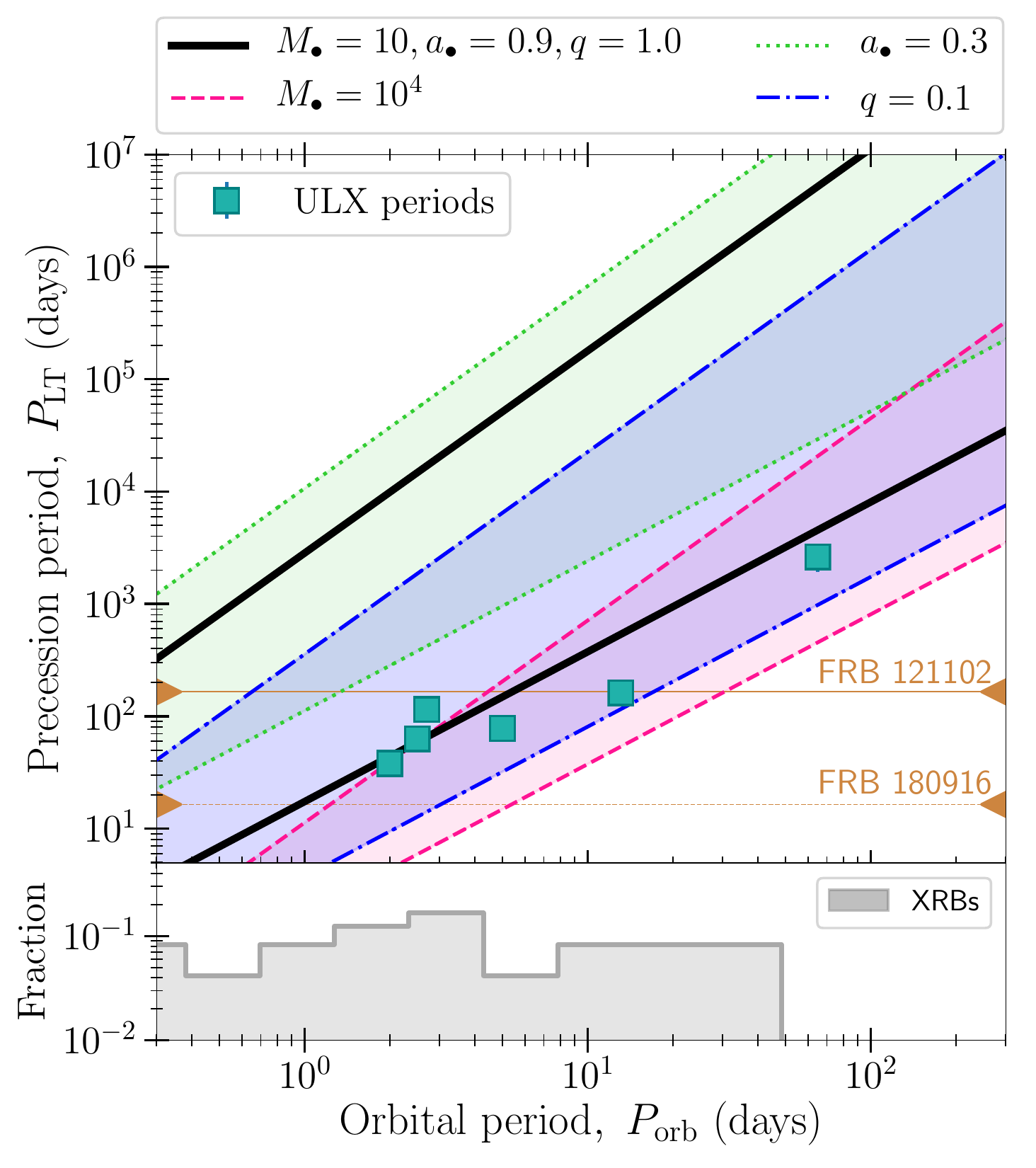}
\caption{\textit{Top panel:} the relationship between the BH-companion star orbital period ($P_{\rm orb}$) and the LT precession period ($P_{\rm LT}$) of an inner thick accretion disk. The band covered between the solid black lines represents the range of $P_{\rm LT}$ values for a fiducial case: $M_\bullet=10M_{\odot},a_\bullet=0.9, q=1.0$. The upper and lower limits of $P_{\rm LT}(P_{\rm orb})$ are set by the nature of the disk outflows---RIAF-like and negligible outflow---parameterized by $p=0.7$ and $p=0$, respectively (see eqs. \ref{eq:ADAF} and \ref{eq:Tprec2}). The region covered between the green (dotted), blue (dashed-dotted), and pink (dashed) bands denotes the independent variation of $a_\bullet (=0.3)$, $q (=0.1)$, and $M_\bullet (=10^4M_{\odot})$, respectively, from the fiducial case.  The observed ${\sim}$160\,d periodicity in FRB 121102 and the ${\sim}$16\,d periodicity in FRB 180916---corresponding to $P_{\rm LT}$ in the paradigm presented here---are denoted by brown horizontal lines connecting the left and right facing triangles. The green squares denote the periodicities observed in ULXs, where we interpret the observed super-orbital periods as $P_{\rm LT}$. The represented ULXs are: NGC 7793 P13 \citep{Motch+14}, SS 433 \citep{Abell&Margon_79}, M82 X-2 \citep{Bachetti+14, Brightman+19}, M51 ULX-7 \citep{Rodriguez-Castillo+20, Brightman+20}, NGC 5907 ULX-1 \citep{Walton+16, Israel+17}, NGC 5408 X-1 \citep{Grise+13}.  \textit{Bottom panel:} the distribution of orbital periods of bright Galactic X-ray binaries ($L_{\rm X}>10^{38}~{\rm erg~s}^{-1}$; gray) obtained from the WATCHDOG catalog \citep{Tetarenko+16}.
}
\label{fig:orb-LT}
\end{figure}

The bursts from FRB 180916 and 121102 arrive in a consistent phase window associated with a period of ${\approx} 16.35 \pm 0.15$\,d and $161 \pm 5$\,d, respectively \citep{CHIME+20b,Rajwade+20,Cruces+21}.  In previous scenarios attributing FRBs to NS activity in a binary, this periodicity was proposed to result from free-free absorption by the companion star wind (e.g., \citealt{Lyutikov+20}).  However, such a scenario predicts a narrower observing window at lower radio frequencies, in tension with observations of FRB 180916  (\citealt{Pastor-Marazuela+20, Pleunis+20}).  By contrast, in the super-Eddington accretion scenario presented here, a more natural explanation arises from geometric beaming by the narrow clean funnel as the BH jet crosses the observer line of sight (Fig.~\ref{fig:cartoon}; see also \citealt{Katz17}).

Two timescales naturally arise in association with a binary, that of orbital motion and that due to disk/jet precession.  An orbital period could manifest in the window of FRB activity in the case of a mildly eccentric binary in which mass-transfer is maximal during pericenter passage.  For a semi-detached binary of mass ratio $q \equiv M_{\star}/M_{\bullet}$, the orbital period is related to the mean density of the mass-transferring star $\bar{\rho}_{\star}$ according to \citep{Paczynski71}
\begin{eqnarray}
P_{\rm orb} &\simeq& 0.35\,{\rm d}\left(\frac{2}{1+q}\right)^{0.2}\left(\frac{\bar{\rho}_{\star}}{\bar{\rho}_\odot}\right)^{-1/2} \nonumber \\
&\underset{q \sim 1}\approx& 0.88\,{\rm d}\,\left(\frac{m_{\star}}{10}\right)^{0.4}\left(\frac{R_{\star}}{R_{\rm \star}^{\rm MS}}\right)^{3/2},
\label{eq:Porb}
\end{eqnarray}
where in the final line we have scaled the stellar radius $R_{\star}$ to its main-sequence value ($r_\star \propto m_\star^{0.6}; \bar{\rho}_{\star} {\approx} m_{\star}/r_{\star}^{3} {\approx} m_{\star}^{-0.8}$).  To reach the observed periods of tens or hundreds of days, the companion star would need to be evolved off the main sequence, consistent with the mass-transfer scenarios outlined in Section \ref{sec:transfer}, for which $R_{\star} {\sim} (1-100)R_{\rm \star}^{\rm MS}$ \citep{Pavlovskii+17}.

\begin{figure}[ht!]
\includegraphics[width=7.8cm]{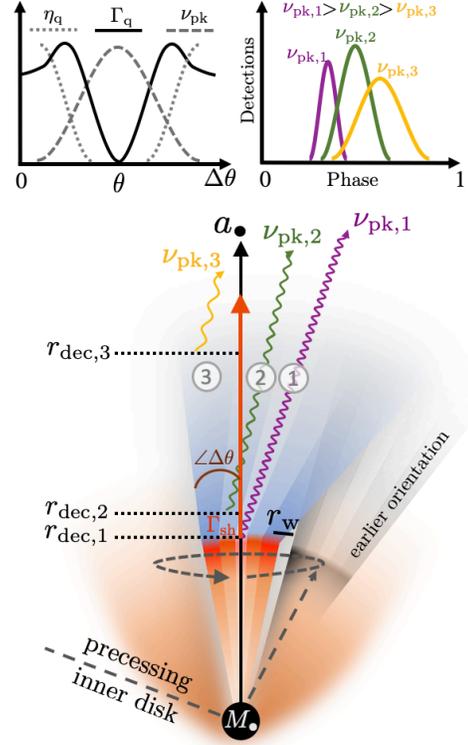}
\caption{A curved quiescent jet cavity (blue bands) is bent toward its earlier orientation (gray cone) due to the drag of the precessing disk winds (brown) at a scale of $r_{\rm w}$ (eq.~\ref{eq:rw}). The FRB-emitting flaring ejecta (red bands) are launched along the instantaneous jet axis (red vertical line with arrow), and propagate into an asymmetric (spirally curved) upstream medium. The radio frequency of the burst depends on whether the flare interacts with the core or sheath of the quiescent jet. The varying properties of the quiescent jet ($\Gamma_{\rm q}$ and $\eta_{\rm q}$; e.g., \citealt{Tchekhovskoy+08}), and the resulting synchrotron maser's peak frequency $\nu_{\rm pk}$ (eq.~\ref{eq:scaling}) as a function of angle $\theta$ from the jet axis, are represented by the schematic at the top-left corner. The larger interaction region of the flare with the sheath of the jet implies that the observed phase window of the low-frequency bursts is wider than that of the higher-frequency bursts. The shock deceleration radius $r_{\rm dec}$ (eq.~\ref{eq:rdec}) is shorter for flares interacting with the spine of the jet compared to the interaction produced near the sheath. This implies that the high-frequency bursts lead the lower-frequency ones, and are shorter in burst width than the lower-frequency ones ($t_{\rm FRB} {\sim} r_{\rm FRB}/(2\Gamma^{2}c)$; \citealt{Metzger+19}). This frequency-dependent phase window, and the arrival times of bursts, are shown by dividing the jet into three regions, and the corresponding detection phase windows (arbitrary normalization) are represented in the schematic at the top-right corner.}
\label{fig:phase_arrival}
\end{figure}

Another source of periodicity can arise due to precession of the accretion funnel along which the FRB is beamed (see section \ref{sec:beaming}, and Figs.\,\ref{fig:cartoon} and \ref{fig:phase_arrival}).  If the spin axis of the accreting BH or NS is misaligned with the angular momentum axis of the disk, then the Lens-Thirring (LT) torque applied by the rotating spacetime on the disk may cause the latter to precess (e.g., \citealt{Middleton+19}).  Numerical simulations have shown that for thick disks (with vertical aspect ratio $h/r {\gtrsim} 0.05$ and low effective $\alpha$ viscosity), the warp propagation timescale is shorter than the differential precession timescale, thereby allowing them to precess as rigid bodies with negligible warping \citep{Fragner&Nelson10}.  The LT precession time in this case is roughly given by (e.g.~\citealt{Fragile+07, Stone&Loeb12})
\be
P_{\rm LT} \approx \frac{\pi GM_{\bullet}}{c^{3}}\frac{(1+2\xi)}{(5-2\xi)}\frac{R_{\rm out}^{5/2-\xi}{R_{\rm in}}^{1/2+\xi}}{R_{\rm g}^{3} a_\bullet},
\label{eq:PLT}
\ee
where $a_{\bullet}$ is the dimensionless BH spin, and $\Sigma \propto r^{-\xi}$ is the surface density of the disk extending from the inner radius $R_{\rm in}{\sim} R_{\rm isco}$\footnote{Recent works by \cite{Sridhar+19, Sridhar+20} and \cite{Connors+20, Connors+21} have tracked the inner accretion flow properties of microquasars across the bright, ballistic jet-emitting states during an X-ray outburst with state-of-the-art X-ray reflection models, and have demonstrated that the inner edge of the accretion disk extends to $R_{\rm in}{\sim} R_{\rm isco}$ at these states.} to an outer radius $R_{\rm out} {\gg} R_{\rm isco}$. For RIAFs, $\Sigma \propto r^{p-1/2}$ (e.g.~\citealt{Blandford&Begelman99}), such that for $p = 0.7$ we have $\xi = -0.2$. On the other hand, $p=0$ (i.e. $\xi=0.5$) when all of the mass-transferred from the companion star reaches the BH ($\dot{M}_{\rm tr}=\dot{M_\bullet}$; see \ref{eq:ADAF}). The precise load of the outflow is uncertain ($0<p<0.7$), and considering its limits, eq.~\ref{eq:PLT} becomes,

\begin{align}
P_{\rm LT} \approx
\left\{\begin{array}{ll}
\frac{\pi}{9a_{\bullet}} \frac{R_{\rm g}}{c}\left(\frac{R_{\rm out}}{R_{\rm g}}\right)^{2.7}\left(\frac{R_{\rm isco}}{R_{\rm g}}\right)^{0.3}, & p=0.7 \\
\frac{7\pi}{8a_{\bullet}} \frac{R_{\rm g}}{c}\left(\frac{R_{\rm out}}{R_{\rm g}}\right)^{2}\left(\frac{R_{\rm isco}}{R_{\rm g}}\right), & p=0.0
\end{array}
\right..
\label{eq:Tprec1}
\end{align}

Taking the outer edge of the disk $R_{\rm out} {\approx} R_{\rm RLOF}/3$ close to the circularization radius \citep{Frank+02}, where (for $0.1 {\lesssim} q {\lesssim} 0.8$) we have \citep{Paczynski71}
\be
\frac{R_{\rm RLOF}}{a} \simeq 0.462 \left(\frac{q}{1+q}\right)^{1/3},
\ee
where $a$ is the binary semi-major axis. Using Kepler's second law, $P_{\rm orb} = 2\pi\left[a^{3}/GM_{\bullet}(1+q)\right]^{1/2}$, eq.~\ref{eq:Tprec1} becomes
\begin{align}
P_{\rm LT} \approx
\left\{\begin{array}{ll}
2.8\times10^3{\rm\, d}\frac{(q_{1.0} P_{\rm orb,d}^2)^{0.9}}{a_{\bullet,0.9}(m_{\bullet}/10)^{0.8}}, & p=0.7 \\
17{\rm\, d}\frac{(q_{1.0} P_{\rm orb,d}^2)^{2/3}}{a_{\bullet,0.9}(m_{\bullet}/10)^{1/3}}, & p=0.0
\end{array}
\right..
\label{eq:Tprec2}
\end{align}

Depending on the BH spin, mass, and mass ratios, precession timescales of tens to thousands of days are possible even from binaries with orbital periods of days (Fig.~\ref{fig:orb-LT}).  A similar model was proposed to explain the 164\,d jet precession timescale of SS433 \citep{Sarazin+80,Katz81}.  Super-orbital periods in the range of tens to hundreds of days, which could be attributed to precession\footnote{Classical mechanisms including Newtonian precession were one of the initial interpretations of the observed super-orbital modulation \citep{Katz_1973, Levine&Jernigan_1982, Katz+82}.}, have been observed in several ULXs (e.g., \citealt{Mioduszewski+04,Grise+13,Atapin&Fabrika16,Luangtip+16,Brightman+19,Brightman+20,Vasilopoulos+20}; see \citealt{Weng&Feng18} for a systematic search with {\it Swift} observatory).

\begin{figure*}
     \begin{subfigure}[b]{0.50\textwidth}
        \includegraphics[width=1.0\textwidth]{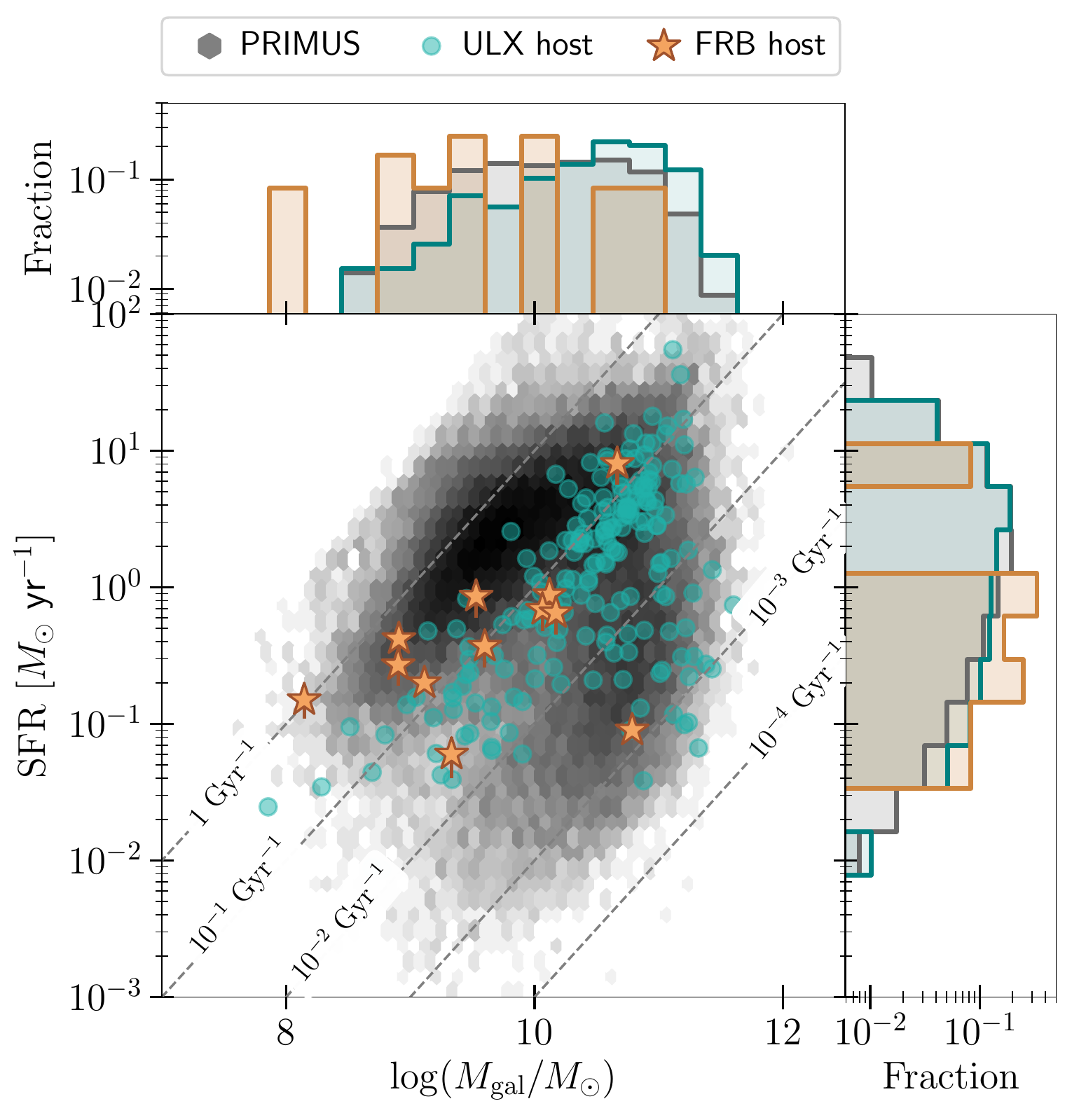}
        \label{fig:SFR-mass}
     \end{subfigure}
     \hfill
     \begin{subfigure}[b]{0.50\textwidth}
        \includegraphics[width=1.0\textwidth]{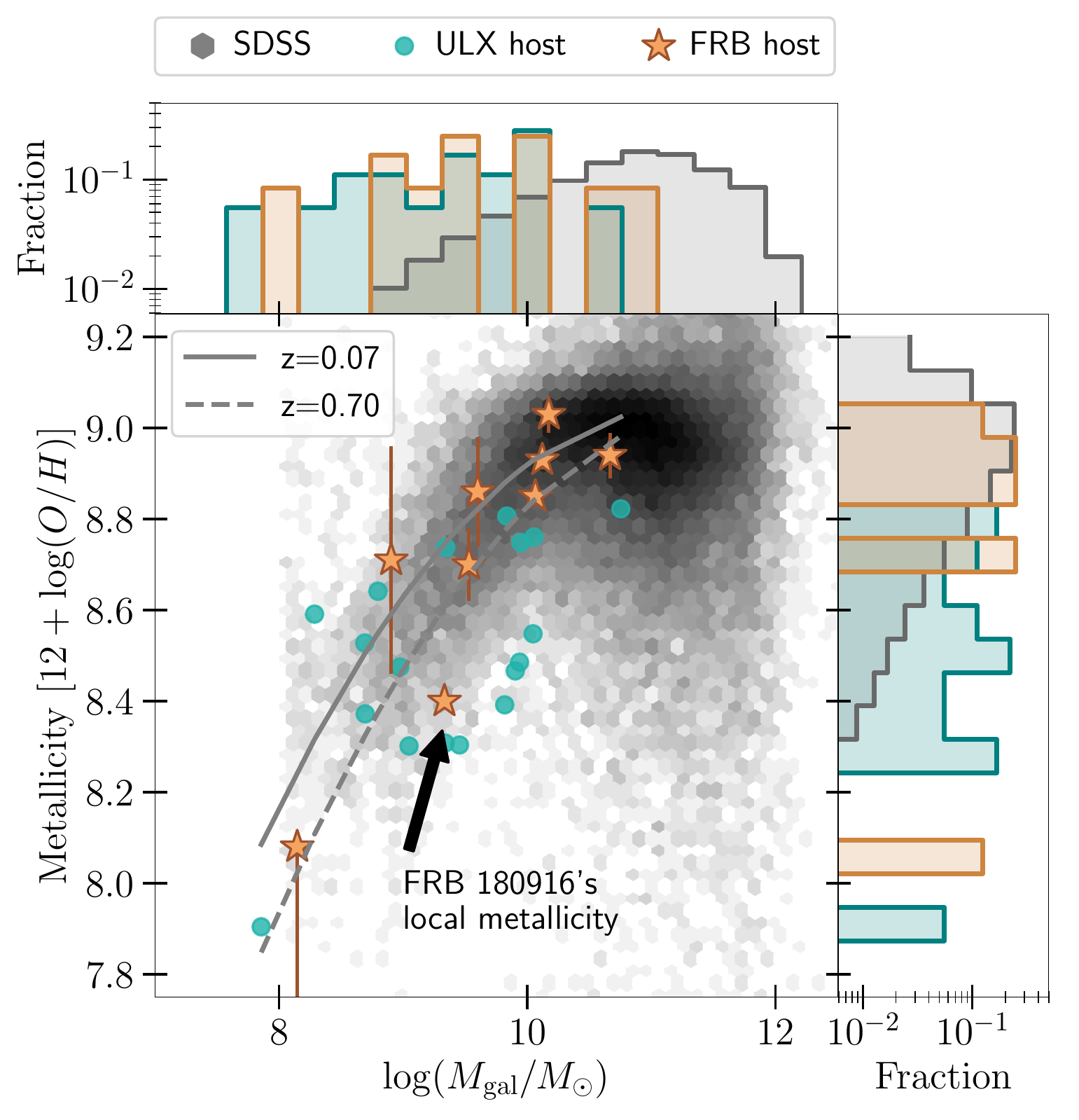}
        \label{fig:SFR-metallicity}
    \end{subfigure}
    \caption{\textit{Left panel:} the stellar mass and the star formation rate (SFR) of the host galaxies of ULXs \citep[green circles;][]{Kovlakas+20} and FRBs \citep[brown stars;][]{Heintz+20}, in comparison to a sample of field galaxies from the PRIMUS catalog \citep[gray;][]{Coil+11}. The gray dashed lines denote the contours of specific SFR. The top and right panels denote the distribution of stellar mass and SFR, respectively; dark gray, brown, and green histograms represent the PRIMUS sample, the FRB hosts, and the ULX hosts, respectively. \textit{Right panel:} the stellar mass and the metallicity of the host galaxies of ULXs and FRBs in comparison to a sample of Sloan Digital Sky Survey (SDSS) emission line star-forming galaxies (gray; e.g., Fig.~9 of \citealt{Heintz+20}). The solid and dashed gray curves denote the empirical mass-metallicity relations \citep{Maiolino+08} for redshifts $z=0.07$ and and $z=0.7$, respectively. The local (${\lesssim}$60\,pc) metallicity of FRB 180916 \citep{Tendulkar+21} is also marked for comparison. The top and right panels denote the distribution of stellar mass and metallicity, respectively; dark gray, green, and brown colors represent the SDSS sample, ULX hosts and FRB hosts, respectively.
    }
    \label{fig:mass-sfr-metallicity}
\end{figure*}

\begin{figure}[ht!]
\includegraphics[width=8cm]{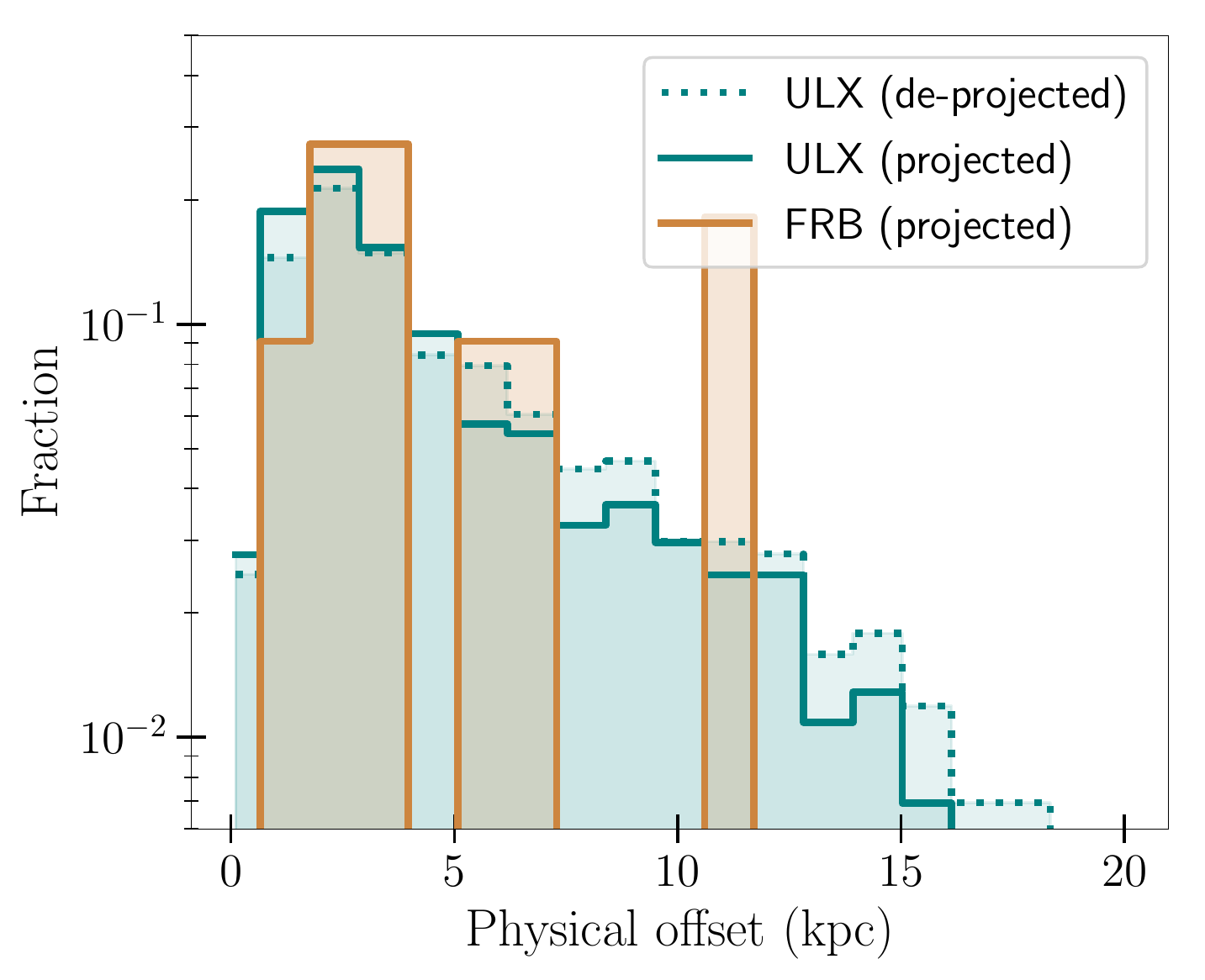}
\caption{The distributions of the projected physical offsets of ULX (solid green) and FRB (brown) sources from their respective host galactic centers are represented by the solid green and brown histograms, respectively. The dotted green histogram corresponds to the de-projected offsets of ULXs, for comparison. FRB offsets are obtained from \citet[][and the references therein]{Heintz+20}, and ULX offsets are calculated from the HECATE-ULX catalog \citep{Kovlakas+20}.
}
\label{fig:offset}
\end{figure}

Observations of the periodic repeater, FRB 180916, reveal that the burst activity window (duty cycle; see eq.~\ref{eq:zeta}) is narrower, and peaks earlier in phase, at higher radio frequencies \citep{Pastor-Marazuela+20}.  Furthermore, the low-frequency bursts are observed exhibit greater average fluences \citep{Pastor-Marazuela+20, Pleunis+20}.  In the synchrotron maser scenario (Section \ref{sec:emission}), the frequency of the radio emission scales with the properties of the FRB-generating transient flare and the pre-existing quiescent jet as (eq.~\ref{eq:nupk}; $L_{\rm f}\propto\eta$),
\be
\nu_{\rm pk} \propto \frac{L_{\rm q}^{3/4}\sigma_{\rm q}^{1/2}}{\Gamma_{\rm q}^{2}L_{\rm f}^{1/4}}.
\label{eq:scaling}
\ee
Thus, if the quiescent jet is ``structured'' in angle $\theta$ measured relative to the jet axis, with both $L_{\rm q}$ and $\Gamma_{\rm q}$ growing toward the jet edge (with $\Gamma_{\rm q}$ growing faster), then lower-frequency bursts would be preferentially observed at phases near the edges of the observing window (see Figs. \ref{fig:cartoon} and \ref{fig:phase_arrival}).  Motivating such a structure, MHD simulations of relativistic magnetized jets find that the jet Poynting flux is concentrated in a hollow cone around the jet core (e.g., \citealt{Tchekhovskoy+08}).  In addition, the efficiency of the synchrotron maser emission depends on the magnetization of the upstream medium \citep{Plotnikov&Sironi19}; thus, angular structure in the magnetization of the quiescent jet $\sigma_{\rm q}(\theta)$ would also imprint systematic variations in burst luminosity across the observing window (and hence with radio frequency; see Fig.~\ref{fig:cartoon}).

Regarding the lag in the central phase of FRB activity with radio frequency \citep{Pastor-Marazuela+20}, we speculate that it results from the curvature of quiescent jet cavity due to the effect of precession-driven motion of the disk winds (see Section \ref{sec:persistent}).  Even if the FRB-emitting flare ejecta is launched ballistically outwards along the axis defined by the instantaneous base of the jet, by the radii of emission ${\gtrsim} r_{\rm dec} {\sim} 10^{14}$ cm (eq.~\ref{eq:rdec}), the quiescent jet medium into which the shock propagates could exhibit asymmetry between the leading and trailing edges of the precession cone, as the wind shaping the jet cavity walls is dragged back against the direction of precession.  Furthermore, the properties of the curved quiescent jet (e.g., $\Gamma_{\rm q}$, $\eta_{\rm q}$) are also expected to be structured about the jet axis \citep{Tchekhovskoy+08}. This can influence the region of interaction of the flare ejecta with the structured quiescent jet, and therefore modulate the emission frequency, burst arrival times, and their activity window. This possibility and how it relates to the angular structure of the jet discussed above is illustrated schematically in Fig.~\ref{fig:phase_arrival}.

Alternatively, some low-frequency FRBs could be bursts of intrinsically higher-frequency viewed slightly off-axis from the direction of the emitting shock front (such that the observed emission is Doppler-shifted to lower frequency).  In this situation, the peak frequency in the observer frame will decrease, and the burst duration increase, both by the same factor \citep{Beniamini&Kumar20}, i.e. $\nu_{\rm off}/\nu_{\rm on}=t_{\rm off}/t_{\rm on}$ (where the subscript off/on corresponds to observed quantities for an observer who is off/on-axis to the direction along the outflow's velocity).  If the relativistic outflow is directed across a narrow range of angles along the core of the jet, this could result in a wider phase window for the low-frequency off-axis bursts.

\section{Environmental Implications}
\label{sec:additional}

\subsection{Host Galaxies}
\label{sec:hosts}

Based on Hubble Space Telescope imaging of eight FRB host galaxies with sub-arcsecond localization, \citet{Mannings+20} fond that FRBs reside in IR-fainter regions, consistent with the locations of core collapse SNe but not of the most massive stars (see also \citealt{Bhandari+20,Heintz+20,Li&Zhang20}).  \citet{Tendulkar+21} likewise ruled out significant star formation or an H\,{\sc ii} region at the location of FRB 180916; their upper limits on the H$\alpha$ luminosity at the burst location constrain potential stellar companions to be cooler than the spectral type O6V.  Given the spatial offset of FRB 180916 from the nearest young stellar clump, \citet{Tendulkar+21} further argued for a source age ${\gtrsim} 0.8-7$ Myr given the expected range of projected velocities of pulsars, magnetars, or NSs in binaries.  These observations are consistent with the scenario described thus far in which FRBs arise from binaries undergoing mass-transfer from a companion star following its main-sequence lifetime, which is ${\sim} 20(10)$ Myr for a $10(20)M_{\odot}$ star \citep[e.g.,][]{zapartas:17}.

\citet{Poutanen+13} find a spatial correlation in the Antennae galaxies between the ULX sources and young stellar clusters (${<}6$ Myr).  Furthermore, they show that most ULXs are displaced outside of the clusters, suggesting that the massive ULX binaries were ejected out of the star clusters---likely due to strong gravitational encounters.  FRBs also show evidence for offsets from regions of intense star formation \citep{Mannings+20,Tendulkar+21} and have in at least one case been localized to a region between two potentially interacting galaxies \citep{Law+20}, perhaps akin to a more distant version of the Antennae.

If FRBs arise from mass-transferring binaries similar to ULX sources, they would be expected to occupy similar host galaxies and locations within their hosts.  The left panel of Fig.~\ref{fig:mass-sfr-metallicity} shows the star formation rate (SFR) and stellar mass ($M_{\rm gal}$) of FRB hosts \citep{Heintz+20} compared to those of ULX hosts \citep{Kovlakas+20}.  The hosts of FRBs and ULXs generally form stars at lower rates than in normal galaxies at a given stellar mass (e.g., below the main sequence of the star formation galaxies, or the locus with a higher specific SFR of ${\sim}1~{\rm Gyr}^{-1}$; see also \citealt{Heintz+20}). While the hosts of some FRBs and ULXs are also in the quiescent cloud (e.g., \citealt{Swartz+11, Walton+11, Ravi_19}), they both occur at a greater frequency in hosts with SFR ${\gtrsim}0.1\,M_{\odot}~{\rm yr}^{-1}$, as seen from the field galaxy population.  \citet{James+21} found that the FRB rate, $\Phi(z) \propto$ (SFR(z)/SFR($z=0$))$^{n}$ evolves with SFR relative to the cosmological average SFR($z$) with a power-law index $n = 1.36_{-0.51}^{+0.25}$.  This is consistent with the SFR-evolution of the ULX population, for which $n = 0.91_{-0.15}^{+0.10}$ \citep{Mapelli+10}.

The FRB hosts are seen to be systematically less massive than those in our ULX sample. As emphasized by \citet{Bochenek+20b}, comparing FRB host galaxies (typical redshift $z {\sim} 0.1-1$) to nearby galaxy populations such as our ULX sample (${\lesssim}$40\,Mpc) can be problematic due to the cosmological evolution effects, particularly the shift in star-formation to lower-mass galaxies with decreasing redshift.  If FRBs indeed trace star formation, this would further exacerbate the tension between ULX and FRB host galaxy masses. On the other hand, the ULX surveys are systematically biased against low-mass galaxies due to selection effects.\footnote{An exact comparison between the HECATE-ULX galaxies and the PRIMUS galaxy sample is also complicated by the different indicators (spectral bands) used to determine their respective SFRs.  For example, infrared-based SFR indicators are most reliable for late-type galaxies.}

The right panel of Fig.~\ref{fig:mass-sfr-metallicity} compares the metallicities of the FRB and ULX hosts in our sample.\footnote{We do not include the ULX samples that host AGNs in our mass-metallicity map. This is because, the oxygen-to-nitrogen emission-line flux ratios are overestimated in AGNs, resulting in an underestimate of the galaxy's metallicity \citep{Nagao+06, Stampoulis+19}}  
ULXs exhibit a clear preference for lower metallicities (e.g., \citealt{Mapelli+10, Walton+11,Prestwich+13}), which theoretically has been attributed to the prevalence of more massive BHs at low metallicity (e.g., \citealt{Heger+03}) and other binary evolutionary effects (e.g.~\citealt{Linden+10,Pavlovskii+17,Marchant+17}). On the other hand, Fig.~\ref{fig:mass-sfr-metallicity} shows that the FRB host population can also extend to lower metallicities than most of the ULX hosts.  At the location of FRB 180916, \citet{Tendulkar+21} found a low metallicity of $12+\log(O/H){\simeq} 8.4$ corresponding to $Z {\approx} Z_{\odot}/2$. Furthermore, the observed ULX population could be biased toward long-lived binary systems, while FRB sources may conversely be produced by binary systems undergoing unstable mass-transfer (i.e., those that reach the highest accretion rates; Section \ref{sec:rates}).  Systems undergoing unstable mass-transfer may preferentially occur at higher-metallicity, e.g. due to the dependence of the stability criterion on the mass of the primary BH.  A fraction of ULXs may also arise from intermediate-mass BHs (\citealt{Miller&Colbert04}), which are potentially too massive to generate the shortest FRBs (Fig.~\ref{fig:t-e}).

The projected physical offsets of FRB and ULX sources from their respective host galaxy centers are represented in Fig.~\ref{fig:offset}. The peaks of the two offset distributions are seen to largely overlap with each other. We note here that our ULX sample is biased against larger offsets\footnote{The ULX sources from the HECATE-ULX sample \citep{Kovlakas+20} are drawn only from within the D25 region of the host galaxy (i.e., the radius of the isophote where the observed band's surface brightness is 25\,mag\,arcsec$^{-2}$). Therefore, the spatial distribution of the ULXs within their hosts is limited by the radius of the selection region, and is devoid of sources with larger offsets belonging to an exponentially decreasing tail of the distribution.}, and the FRBs can exhibit a deficit of short offset sources. The latter may not be an intrinsic effect.  The population of FRB sources with small offsets may be underestimated, with decreasing FRB detections from regions closer to the host center. This could be attributed to the increased smearing of the signal by the denser interstellar medium, accompanied by an increase in the DM (e.g., \citealt{Heintz+20}).

\subsection{Local Environment and Nebular Emission}
\label{sec:persistent}

\begin{figure}[ht!]
\includegraphics[width=8cm]{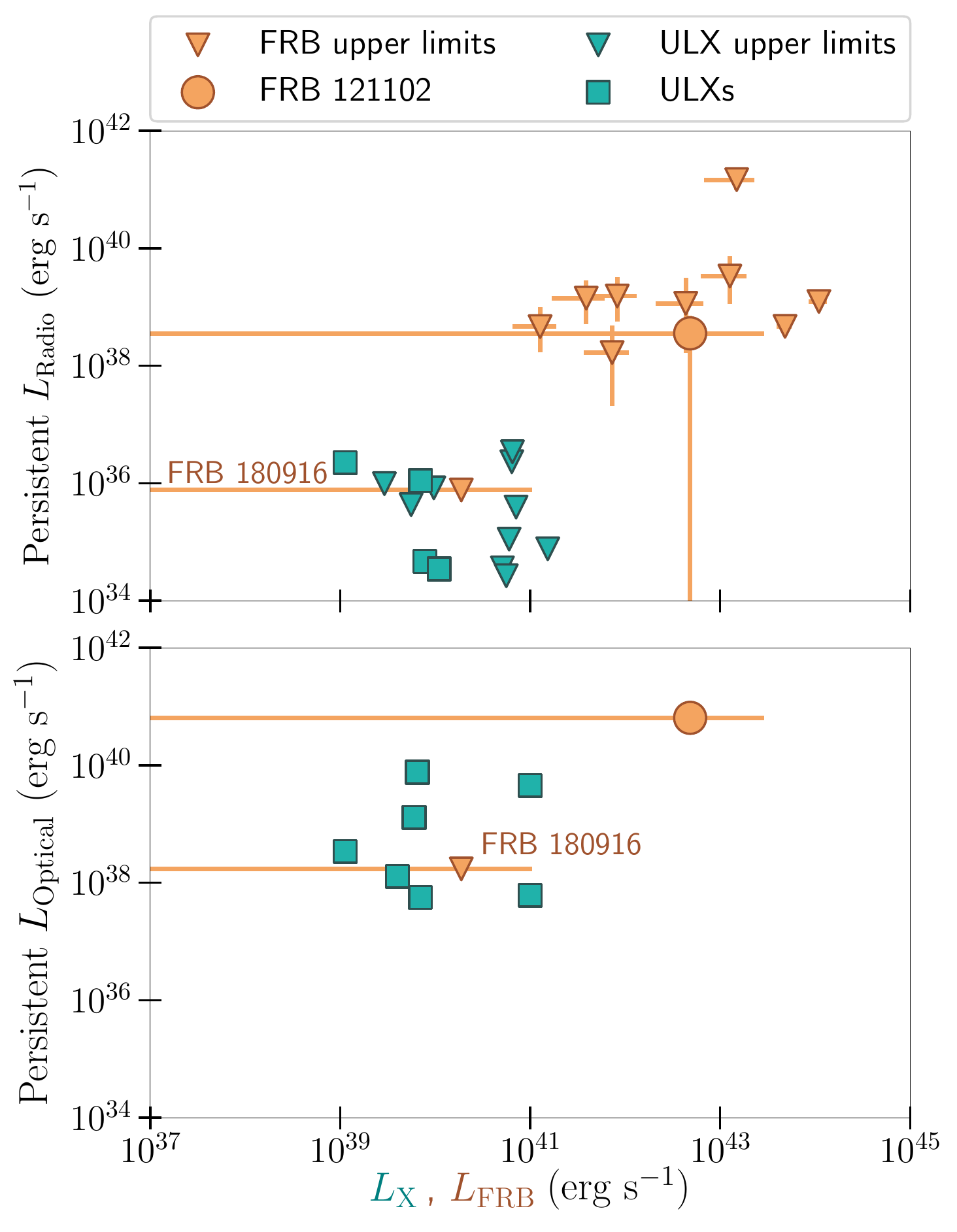}
\caption{
Luminosities of the persistent radio (upper panel) and optical (lower panel) counterparts to FRBs (brown) and ULXs (green) against their respective radio burst or X-ray luminosities (horizontal axis). Detections and upper limits are shown with squares and upside down triangles, respectively.  Persistent radio counterparts and the X-ray luminosities of the represented ULXs are obtained from \citet{Kaaret+03}, \citet{Miller+05}, \citet{Roberts+06}, \citet{Mezcua+13a, Mezcua+13b}, \citet{Sutton+13a, Sutton+13b}, \citet{Luangtip+16}, and \citet{Earnshaw+19}. The persistent FRB radio luminosities are obtained from \cite{Yang+20} (and the references therein), and the ULX persistent optical counterparts are calculated from the reported optical magnitudes for Holmberg IX X-1, NGC 5204 X-1 \citep{Pakull&Mirioni02}, Holmberg II X-1 \citep{Lehmann+05}, HLX-1 \citep{Soria+10}, NGC 6946 \citep{Kaaret+10}, NGC 4559 X-10, and NGC 4395 ULX-1 \citep{Vinokurov+18}.  The isotropic-equivalent luminosity of the persistent optical counterparts to FRB 121102 and 180916 are calculated from the optical magnitudes reported in \citet{Chatterjee+17} and \citet{Tendulkar+21}, respectively.
}
\label{fig:persistent}
\end{figure}

A hallmark of super-Eddington accretion flows is mass-loaded outflows (e.g., \citealt{Blandford&Begelman99,King+01}). ULX systems exhibit direct evidence for winds with super-Eddington mechanical powers \citep{Soria+14} carrying a total energy ${\sim} 10^{52}$\,erg \citep{Roberts+03}, which can exceed the radiative output by up to a factor of 1000 \citep{Pakull+10}.

The mass-loss rate in super-Eddington winds $\dot{M}_{\rm w}$ is comparable to the binary mass-transfer rate $\dot{M}_{\rm tr}$ (eq.~\ref{eq:Mdottr}).  The density profile of such an outflow of velocity $v_{\rm w}{\sim}0.1c$ \citep{Pinto+16}, at radii much larger than the binary separation, can be approximated as
\be
n_{\rm w} \simeq \frac{\dot{M}_{\rm w}}{4\pi v_{\rm w} m_{\rm p} r^{2}}.
\ee

If the accretion disk undergoes precession with a period $P$ (Fig.~\ref{fig:cartoon}), then any ballistic ejection of material along the instantaneous jet direction will encounter the disk wind material from an earlier orientation of the disks by a radial scale no larger than:
\be
r_{\rm w} \sim P\zeta v_{\rm w} \approx 7.5\times 10^{15}{\rm cm}\left(\frac{\zeta}{0.3}\right)\left(\frac{P}{100\,{\rm d}}\right)\left(\frac{v_{\rm w}}{0.1 c}\right).
\label{eq:rw}
\ee
The Thomson column of this material encountered by an FRB emitted at smaller radii can be estimated as
\begin{eqnarray}
&&\tau_{\rm T,w} \simeq n_{\rm w}(r_{\rm w})\sigma_{\rm T} r_{\rm w} \approx \frac{\dot{M}_{\rm w} \sigma_{\rm T}}{4\pi m_{\rm p} v_{\rm w} r_{\rm w}} \nonumber \\
&\sim& 1.8\times 10^{-9}\left(\frac{\dot{M}_{\rm w}}{\dot{M}_{\rm Edd}}\right)\left(\frac{m_{\bullet}}{10}\right)\left(\frac{r_{\rm w}}{10^{16}{\rm cm}}\right)^{-1}\left(\frac{v_{\rm w}}{0.1\,c}\right)^{-1}.
\end{eqnarray}
For $\dot{M}_{\rm w} {\gg} \dot{M}_{\rm Edd}$, time variability of this column could generate variations in $DM {\approx} 5\times 10^{5}\tau_{\rm T}$ pc cm$^{-3}$ over the timescale $P$, distinct from that accumulated through the quiescent jet cavity (eq.~\ref{eq:tauT}). If the accretion disk wind is magnetized, then propagation through it could generate a more significant local time-variable RM across the active phase (see inset of Fig.~\ref{fig:cartoon}), as observed in FRB 180916 \citep{Pleunis+20} and FRB 121102 (\citealt{Hessels+19}; although the latter could originate from an AGN or a wind-fed nebula on larger scales; \citealt{Margalit&Metzger18}).  As with the DM above, we estimate the RM contribution through the precessing disk wind (at radii ${\gtrsim} r_{\rm w}$) to be
\begin{eqnarray} \label{eq:RM}
&&{\rm RM} \sim \frac{e^{3}}{2\pi m_{\rm e}^{2}c^{4}}B_{\rm w}n_{\rm w}(r_{\rm w})r_{\rm w}\sin \alpha \approx 0.5\,{\rm rad\,m^{-2}}\left(\frac{\sin \alpha}{0.1}\right) \times \nonumber \\
&&\left(\frac{\sigma_{\rm w}}{0.01}\right)^{1/2} \left(\frac{\dot{M}_{\rm w}}{\dot{M}_{\rm Edd}}\right)^{3/2}\left(\frac{m_{\bullet}}{10}\right)^{3/2}\left(\frac{v_{\rm w}}{0.1\,c}\right)^{-3/2}\left(\frac{r_{\rm w}}{10^{16}{\rm cm}}\right)^{-2},
\end{eqnarray}
where $\sigma_{\rm w} = B_{\rm w}^{2}/(4\pi n_{\rm w} m_{\rm p} c^{2})$ is the radially constant magnetization of the wind, and $\alpha$ is an appropriate average angle between the directions of the magnetic field and the FRB path of propagation.

At larger radii $r {\sim} 10^{17}$ cm, the wind density for fiducial parameters ($\dot{M}_{\rm w} {\sim} 10^{3}\dot{M}_{\rm Edd}$; $M_{\bullet} {\sim} 10M_{\odot}$; $v_{\rm w} {\lesssim} 0.1c$) is $n_{\rm w} {\gtrsim} 10$ cm$^{-3}$.  This provides the required plasma for self-modulation of the FRB signal \citep{Sobacchi+21}, which predicts the duration of the sub-burst structures to be ${\sim}30\mu{\rm s}/\sqrt{n_{\rm w,1}}$.

On yet larger scales of up to hundreds of parsecs, ULX outflows are observed to inflate bubbles of shock-ionized plasma capable of generating persistent optical, X-ray, and radio emission \citep{Pakull&Mirioni02,Wang02,Kaaret+03,Ramsey+06,Soria+10}.  The ULX jet also powers steady non-thermal synchrotron emission on smaller scales \citep{Miller+05,Lang+07}.

Fig.~\ref{fig:persistent} compiles detections and upper limits on the persistent radio (top panel) and optical (bottom panel) luminosities of FRBs and ULXs.  Although many ULXs are detected, most of the upper limits on FRB persistent source emission are unconstraining due to their comparatively greater distances.  Nevertheless, if some FRBs indeed arise from long-lived ULX sources, the closest events may eventually exhibit detectable, and potentially resolvable, persistent emission.  This would not necessarily be true for the subset of FRBs arising from short-lived binaries undergoing unstable mass-transfer (Section \ref{sec:binary}), if the system is not active long enough to inflate a large bubble.  

The compact synchrotron radio source co-located to $< 0.7$ pc of FRB 121102 \citep{Chatterjee+17,Marcote+17} could arise from a young (${\sim}$decades old) ULX-bubble from a binary system in the process of undergoing unstable mass-transfer.  The total energy ${\sim} 10^{50}-10^{51}$ erg and mass-outflow rate $\dot{M} {\sim} 10^{19}-10^{21}$ g s$^{-1}$ needed to simultaneously explain the radio spectrum and RM of the source (\citealt{Margalit&Metzger18}), are consistent with those expected from super-Eddington disk outflows.

Furthermore, persistent X-ray emission is also expected from FRBs that are produced via the ULX channel.  However, due to the large distances of the hitherto detected cosmological FRBs, persistent X-ray emission at the luminosities of ULX sources cannot be detected. The currently tightest constraint indeed comes from one of the nearest repeaters FRB 180916 (at ${\sim}$150\,Mpc), for which the 0.5--10 keV is constrained to be $L_{\rm X} {\lesssim} 2\times 10^{40}$ erg s$^{-1}$ \citep{Scholz+20}, corresponding to ${\sim} 10-100 L_{\rm Edd}$ for a BH or NS accretor, respectively. On the other hand, the relatively fainter bursts (${\sim}10^{37}$erg~s$^{-1}$; see Fig.~\ref{fig:t-e}) from the nearest observed repeating FRB 200120 (at ${\sim}$3.6\,Mpc), can be predicted to exhibit an X-ray flux of ${\sim}10^{-13}$erg~s$^{-1}$cm$^{-2}$ through the accretion channel. This is well below the sensitivity limit of existing X-ray telescopes, consistent with the reported X-ray upper limits from the source error region \citep{Bhardwaj+21}.

\subsection{Multiwavelength Transient Counterparts}
\label{sec:counterpart}

Models in which FRBs are generated by magnetized shocks predict simultaneous electromagnetic emission ranging from the optical to gamma-ray bands arising from (incoherent) thermal synchrotron emission (e.g.,~\citealt{Metzger+19,Beloborodov19}).  Though carrying more radiated energy than the FRB itself, this counterpart is challenging to detect except for the brightest FRB sources, such as those originating from Galactic magnetar flares.  X-rays were observed simultaneously with the FRB-like burst from SGR 1935+2154 (e.g., \citealt{Mereghetti+20}), consistent with the predictions of shock synchrotron emission \citep{Metzger+19,Margalit+20}.  

ULXs are known to exhibit fluctuating non-thermal mid-IR outbursts on timescales of a few days that are attributed to their variable jets \citep[e.g., Holmberg IX X-1;][]{Lau+19}. These events can be related to traversing magnetized shocks in precessing jets---that are expected to give rise to an active window of FRBs. Therefore, we strongly encourage simultaneous X-ray and IR monitoring of nearby FRB sources (a few megaparsecs)---especially during the FRB active window, in order to discern their connection to jetted accreting sources.

As discussed in Sections \ref{sec:transfer} and \ref{sec:rates}, the most luminous accretion-powered FRB sources would arise from binary systems undergoing unstable mass-transfer. These systems are predicted to be short-lived, with a lifetime of peak FRB activity of $\tau_{\rm unst} {\lesssim}$ years to decades (eq.~\ref{eq:trun}).  The end state of this process is a stellar merger or common envelope event \citep{Ivanova+13}.

Stellar merger events are commonly observed in the Milky Way and nearby galaxies as optical \citep{Soker&Tylenda06} and/or dusty infrared \citep{Kasliwal+17} transients. Most such systems are believed to arise from the merger of binaries consisting of main-sequence or moderately evolved stars (e.g., \citealt{Tylenda+11}), i.e. not containing BH or NS primaries.  However, if similar transients are generated from BH/NS merger events (as suggested by numerical simulations; e.g., \citealt{Law-Smith+20}), then some FRB sources---particularly the most luminous ones---could be accompanied by an LRN in the months to years after ``turning off'' as FRB sources.  More speculatively, the very high BH accretion rates in such events---which are broadly similar to those achieved in tidal disruption events by stellar-mass BHs---could power luminous optical/X-ray transients (e.g., \citealt{Perets+16,Kremer+21}), perhaps similar to the ``fast blue optical transients" (FBOTs; e.g., \citealt{Drout+14,Margutti+19}).  FRBs could then serve as unique sign posts to this special type of stellar merger event.

LRNe reach peak optical luminosities in the range ${\sim} 10^{39}-10^{41}$ erg s$^{-1}$ (e.g.~\citealt{Pastorello+19}), corresponding to visual apparent magnitudes ${\gtrsim} 21-22$, at the distance of even the nearest FRB 180916 (${\approx} 150$ Mpc).  Such transient emission would be challenging to detect with current optical time-domain facilities, but would make promising targets for future surveys including those conducted with the Vera Rubin Observatory (e.g.~\citealt{LSST+09}).  On the other hand, FBOTs reach peak optical and X-ray luminosities up to ${\sim} 10^{44}$ erg s$^{-1}$, though they maintain this luminoisty only for a few days (e.g., \citealt{Margutti+19}).  We accordingly recommend that a deep optical/IR and X-ray follow-up campaign be targeted on previously periodic FRB sources which suddenly ``turn off'' with a ${\lesssim}$ weekly cadence to capture the brightest phases of putative LRN/FBOT emission. 

If the BH spiraling inwards during the common envelope phase generates a relativistic jet, these events could also be a source of high-energy neutrino emission \citep{Grichener&Soker21}.  The end state of such mergers is uncertain, with possibilities ranging from an isolated compact binary with a white dwarf secondary to a quasi-spherical Thorne-Zytkow object \citep{Thorne&Zytkow75}.

\section{Predictions and Conclusions}
\label{sec:conclusions}

We have outlined a scenario in which recurring FRBs are powered by transient flares from accreting stellar-mass BH or NS binary systems. The required high accretion rates, which must exceed the Eddington rate to explain the most luminous FRBs, drive us to consider a potential connection to ULX sources, the closest known class of persistent super-Eddington sources.  We have provided semi-quantitative arguments showing how such systems could in principle account for the observed durations, energetics, beaming fraction, radio frequencies, periodic behavior, rates, and host galaxy properties (with some important caveats).  Each of these issues merits more detailed follow-up studies.

We conclude by enumerating a few implications and predictions of the ULX binary scenario.

\begin{itemize}

\item  Systems with higher accretion rates generate FRBs with narrower beaming fractions if the latter are shaped by the geometry of the super-Eddington accretion flow (e.g., $f_{\rm b} \propto \dot{M}_{\bullet}^{-2}$; eq.~\ref{eq:fb}).  For periodic sources, this could manifest as a narrower active phase with increasing isotropic-equivalent FRB luminosity, i.e., the observed average engine power output (product of burst rate and isotropic-equivalent burst energy) is independent of $f_{\rm b}$, for a beam (with a constant opening angle) distributed uniformly in time about the accretion funnel.  On the other hand, coupled with the potentially shorter lifetimes of high-$\dot{M}_{\bullet}$ systems, this results in a lower probability of identifying the most luminous FRB sources as recurring in the first place.

\item  The properties of FRB host galaxies, and the spatial offsets of the bursting sources within these galaxies, could track those of the most luminous ULX sources (Fig.~\ref{fig:mass-sfr-metallicity}).  However, this correspondence may not be perfect, for instance if FRBs preferentially arise from short-lived binaries undergoing unstable mass-transfer (and hence are biased against being discovered as persistent ULXs).  On smaller spatial scales, some FRB sources will coincide with luminous optical line, X-ray, radio emitting regions akin to the super-bubbles observed surrounding luminous ULXs (Fig.~\ref{fig:persistent}).

\item A fraction of known ULXs may emit detectable FRBs, and we encourage systematic radio monitoring of these sources.  We caution that a rate comparison suggests that atypical conditions---such as a highly magnetized secondary star---may be necessary for ULXs to generate the bulk of the luminous FRBs detected at cosmological distances (Section \ref{sec:rates}).  Nevertheless, weaker radio bursts could be detected from a targeted ULX search due to their comparatively closer distances.  Evidence for transient ultra-relativistic $\Gamma > 100$ outflows from these systems would also support an FRB connection (Fig.~\ref{fig:quiescent}).

\item  Periodic sources could exhibit small systematic variations in the burst DM and RM across the active phase window due to the bursts propagating through the magnetized super-Eddington disk outflows (swept into a spiral pattern intersecting the instantaneous jet axis due to precession of the disk angular momentum).  If the quiescent jet is sufficiently dilute, or the disk precession period sufficiently short, the FRB-generating flare will interact with the disk wind, potentially giving rise to a more complicated evolution of the polarization across the burst than in cases in which the upstream medium is a comparatively organized quiescent jet.

\item The most luminous FRB sources could arise from binaries undergoing unstable mass-transfer and evolving toward a merger or common envelope event.  To the extent that such sources are periodic, this would lead to systematic variations in the average burst properties with time as the accretion rate rises and the environment surrounding the FRB source evolves toward the final dynamical plunge. Such short-lived sources would not exhibit the large several-parsec-scale nebul\ae{} but they could generate compact radio nebul\ae{}, perhaps similar to that observed around FRB 121102.

The unstable accreting sources will ``turn off'' as FRB emitters on a timescale of weeks to years approaching the dynamical phase of the merger.  Soon after this point, the system will generate a luminous optical/IR transient, akin to an LRN or dusty infrared transient.  More speculatively, the very high BH accretion powers in these systems could power much more luminous optical/X-ray emission, perhaps similar to the observed FBOT transients.

\item  Observed FRB durations are consistent with arising from stellar-mass compact objects.  However, a similar physical model could in principle be extended to more massive, intermediate-mass or even super-massive BHs.  Scaling to the larger ISCO radius (minimum variability time), one would predict the existence of ``slow radio bursts'' (SRBs) with larger maximum energy and durations significantly longer than hundreds of milliseconds (see \citealt{Zhang20} for a different physical mechanism for generating SRBs).  This possibility will be the focus of future work.

\end{itemize}

Following the submission of this work, a paper by \citet{Deng+21} appeared on a related FRB model invoking transient relativistic ejections from accreting compact objects.  Although the details of our models differ, their conclusions broadly mirror those presented here.

\acknowledgements

This work benefited from the valuable comments from and discussions with Shreya Anand, Wen-fai Fong, Kasper Heintz, Jason Hessels, Phil Kaaret, Jonathan Katz, James Miller-Jones, Douglas Swartz, Shriharsh Tendulkar and Georgios Vasilopoulos.

N.S. and L.S. are supported by NASA Astrophysics Theory Program (ATP) 80NSSC18K1104 and NSF AST-1716567. B.D.M. acknowledges support from the NASA ATP (grant No. NNX17AK43G), Fermi Guest Investigator Program (grant No. GG016287), and the NSF through the AAG Program (grant No. GG016244). The research of P.B. was funded by the Gordon and Betty Moore Foundation through grant GBMF5076. K.K. acknowledges support from the the Swiss National Science Foundation Professorship grant (project No. PP00P2 176868). K.K. received funding from the European Research Council under the European Union's Seventh  Framework Programme (FP/2007-2013) / ERC Grant Agreement No. 617001.

\software{Astropy \citep{astropy:2013, astropy:2018}, SciPy \citep{2020SciPy-NMeth}}.

\bibliography{references}{}
\bibliographystyle{aasjournal}

\end{document}